\documentclass[aip,jcp,twocolumn,reprint,showpacs,superscriptaddress,floatfix]{revtex4-1}
\usepackage{amstext,amssymb}
\usepackage{amsmath}
\usepackage{indentfirst}
\usepackage{makeidx}
\usepackage{slashed}
\usepackage{graphicx}
\usepackage{xcolor}
\usepackage{dsfont}
\usepackage{adjustbox}
\newcommand{\leftdot}[1]{\!\!\dot{\,\;{#1}}}
\newcommand{\sm}{\scriptscriptstyle}
\newcommand{\h}{H_{{\sm 0}}}
\newcommand{\rhI}{\rho(t_{{\sm 0}})}
\newcommand{\iu}{i}
\newcommand{\Sc}{\mathbb{S}}
\newcommand{\He}{\mathbb{H}}
\newcommand{\In}{\mathbb{I}}
\newcommand{\wz}{\omega_{{\sm 0}}}
\newcommand{\fr}{\Omega}

\begin{document}
\title{Improved Dyson series expansion for steady-state quantum transport beyond the weak coupling limit - divergences and resolution}
\author{Juzar Thingna}
\affiliation{Institute of Physics, University of Augsburg, Universit\"atsstrasse 1 D-86135 Augsburg, Germany}
\affiliation{Nanosystems Initiative Munich, Schellingrstrasse 4, D-80799 M\"unchen, Germany}
\author{Hangbo Zhou}
\affiliation{Department of Physics and Centre for Computational Science and Engineering, National University of Singapore, Singapore 117551, Republic of Singapore}
\affiliation{NUS Graduate School for Integrative Sciences and Engineering, National University of Singapore, Singapore 117456, Republic of Singapore}
\author{Jian-Sheng Wang}
\email[]{phywjs@nus.edu.sg}
\affiliation{Department of Physics and Centre for Computational Science and Engineering, National University of Singapore, Singapore 117551, Republic of Singapore}

\date{\today}

\begin{abstract}
We present a general theory to calculate the steady-state heat and electronic currents for nonlinear systems using a perturbative expansion in the system-bath coupling. We explicitly demonstrate that using the truncated Dyson-series leads to divergences in the steady-state limit, thus making it impossible to be used for actual applications. In order to resolve the divergences we propose a unique choice of initial condition for the reduced density matrix, which removes the divergences at each order. Our approach not only allows us to use the truncated Dyson-series, with a reasonable choice of initial condition, but also gives the expected result that the steady-state solutions should be independent of initial preparations. Using our improved Dyson series we evaluate the heat and electronic currents upto fourth-order in system-bath coupling, a considerable improvement over the standard quantum master equation techniques. We then numerically corroborate our theory  for archetypal settings of linear systems using the exact nonequilibrium Green's function approach. Lastly, to demonstrate the advantage of our approach we deal with the nonlinear spin-boson model to evaluate heat current upto fourth-order and find signatures of cotunnelling process.
\end{abstract}
\maketitle

\section{Introduction}
\label{sec:1}
Understanding nonequilibrium transport properties of quantum models, with nonlinear interactions, is a formidable task encompassing the fields of physics \cite{dubirmp11, lirmp12}, chemistry \cite{segaljcp03, leitnerjpcb13}, and biology \cite{blankenship02, engelnat07}. Traditionally, the density operator techniques using the quantum master equation (QME) formulation are most suited to study models with nonlinear interactions \cite{segalprl05, segalprb06, espositojpcc10, velizhaninjcp10, juzar-prb12}. Despite their obvious success the QMEs are limited to the weak coupling regime \cite{segalprl05, segalprb06, juzar-prb12} and cannot capture the effects of moderate to strong system-bath interaction. 

To overcome this drawback several numerical techniques based on path integral formulation \cite{Muhlbacher2008, Segal2010, Hutzen2012} and diagrammatic quantum Monte-Carlo \cite{Schiro2009, Schiro2010, Werner2009, Gull2010, Gullprb11, Cohenprb11, Cohennjp13} have been developed. These approaches are excellent to capture the transient behaviour of currents but, since numerical errors increase with time \cite{Schiro2009, Segal2010}, they can only be employed for systems which relax quickly to the steady state \cite{Gull2010, Gullprb11, Cohenprb11, Cohennjp13}. The real-time diagrammatic transport theory  \cite{Schoeller94, Schoellerphysica94, Konig96, Saptsov12, kollerprb10} is also a lucrative method to deal with nonlinear interactions through partial resummation, which in principle could be exact. However, the task of identifying all diagrams is a formidable bottleneck in this approach for practical applications. Another powerful non-perturbative technique applicable to Ornstein-Uhlenbeck like correlations is the hierarchy equation of motion approach \cite{sakurai13jpsj, sakurai14njp} (HEOM). The HEOM formalism can capture the transient and steady states accurately, but in practice can not deal with large system-Hilbert spaces comprising of 100's of levels.

In this work, we take the usual point of view of open quantum systems by considering a composite system consisting of two baths (minimal transport set-up) and a finite system with couplings between them. We then develop a general formulation to evaluate the steady-state heat and electronic currents as a perturbation expansion in the system-bath coupling, treating the nonlinearity exactly. Interestingly, we show that if one approaches this problem using a truncated Dyson series \cite{Fetter71} one encounters divergences in the steady state at all orders of coupling strength, except the lowest \cite{kollerprb10}. 

The divergence in the truncated steady-state Dyson series is strongly rooted in the choice of initial condition $\rho(t_{{\sm 0}})$ for the reduced density matrix. Typically $\rho(t_{{\sm 0}})$ is an arbitrary choice as long as one sums up the \emph{entire} Dyson series. Unfortunately, for all practical purposes truncation is unavoidable and in such scenarios we argue that one must choose $\rho(t_{{\sm 0}})$ carefully to avoid divergences. Specifically, the initial condition must be chosen such that it leads to the steady state in long, but \emph{finite} time. This prudent choice of initial condition would then no longer be arbitrary since a unique one-to-one map does exist between the steady state reduced density matrix $\rho_{{\sm SS}}(\mathcal{T})$ and the initial condition $\rho(t_{{\sm 0}})$, i.e., $\rho(t_{{\sm 0}}) = \mathcal{K}^{-1}(t_{{\sm 0}},\mathcal{T})\left[\rho_{{\sm SS}}(\mathcal{T})\right]$, where $\mathcal{T}$ is some large but finite time. Naturally in order to ensure that $\rho_{{\sm SS}}$ is the correct steady-state reduced density matrix one requires an additional constraint of $d\rho_{{\sm SS}}(t)/dt|_{t=\mathcal{T}} = 0$. 

Our choice of explicitly evaluating the unique initial reduced density matrix $\rho(t_{{\sm 0}})$ through a set of equations eliminates the divergences in the truncated-Dyson series expansion. This then allows us to evaluate currents at higher orders in system-bath coupling, which is a considerable improvement over the existing weak-coupling QME techniques. In order to corroborate our theory, we first compare with the exactly solvable systems upto fourth-order in coupling strength. The two generic set-ups we consider involve bosons and fermions as carriers. In the bosonic case we consider an oscillator connected to harmonic baths at two different temperatures and in the fermionic case we consider a spinless fermionic system connected to baths comprising of non-interacting fermions at different chemical potentials. Currents in both these cases can be evaluated exactly using either the nonequilibrium Green's function formalism (NEGF) \cite{wang08review, wang13review, diventra08} or the Langevin equation approach \cite{dhar-roy-jsp-06, dharrev08, tanimurajpsj06}. The strength of our technique lies in its ability to deal with nonlinear interactions and hence we consider the simplest nonlinear model of a spin connected to bosonic baths, also commonly known as the spin-boson model \cite{leggett87review, weiss08}. The spin-boson model has been the topic of intense theoretical research \cite{segalprl05, segaljcp05, nanjcp09, ruokolaprb11}, mainly due to its simplicity and its unusual properties in the strong coupling regime \cite{saitoprl13, segalpre14}. 

The outline of the paper is as follows. In Sec.\ref{sec:2} we present some basic preliminaries and the models considered in this work. We then go on to discuss the Dyson expansion and its shortcomings in Sec.\ref{sec:3}. In Sec.\ref{sec:4} we outline our theory to uniquely fix the initial condition and discuss its connections with the general time-local QME. Then we go on to discuss the fourth-order formulas for currents in Sec.\ref{sec:5} and give a lucid diagrammatic approach to our method. Section \ref{sec:6} is dedicated to illustrative examples and corroborations of our approach upto $4$th order in coupling strength. Finally, we end with some concluding remarks in Sec.\ref{sec:7}. 

\section{Definitions and Models}
\label{sec:2}
We start by introducing the three pictures of quantum mechanics -- the Schr{\"o}dinger, Heisenberg, and the interaction (Dirac) picture. The operators in these pictures will be denoted by $O_{{\sm \Sc}}$, $O_{{\sm \He}}(t)$, and $O_{{\sm \In}}(t)$ respectively. In the Schr{\"o}dinger picture operators do not depend on time, if they do it is an explicit time-dependence (such as an alternating electric field or an adiabatic switch-on parameter). The total Hamiltonian of the combined system, baths, and their mutual interaction is denoted by $H(t)=\h + e^{\epsilon t}V$, where $\h=H_{{\sm S}} + H_{{\sm L}} + H_{{\sm R}}$ is the decoupled Hamiltonian including the system, and the left and the right baths. We will make no particular assumptions on the system Hamiltionian $H_{{\sm S}}$ and the coupling $V$ until the calculation stage at the end. The explicit time dependence $e^{\epsilon t}$ ($\epsilon >  0$) is due to the adiabatic switch-on from the remote past. The choice of the adiabatic switch-on clearly implies that at $t=-\infty$ (initial time) the system and the baths are decoupled. The system and the coupling are then fully turned on at time $t=0$ (final time) and we set the coincidence time for all three pictures at the final time \cite{Fetter71}.

The relations among the pictures are given by unitary transformations, e.g., the Schr{\"o}dinger picture evolution operator $U(t,t') = \mathrm{T} \exp\bigl[-(\iu /\hbar) \int_{t'}^t H(t'') dt''\bigr]$, $t \geq t'$, where $\mathrm{T}$ is the standard time-ordering super-operator. $U_{{\sm 0}}(t,t')$ is similarly defined except that it is associated with the decoupled Hamiltonian $\h$.  Then the unitary transformations relating the Schr\"odinger and the Heisenberg picture are given by
\begin{eqnarray}
O_{{\sm \He}}(t) &=& U(0,t) O_{{\sm \Sc}} U(t, 0), \nonumber \\
\hat \rho_{{\sm \He}}(t) &=& U(0,t) \hat \rho_{{\sm \Sc}}(t) U(t,0). 
\end{eqnarray}
Equivalently the interaction picture and the Schr\"odinger picture are related by
\begin{eqnarray}
O_{{\sm \In}}(t) &=& U_{{\sm 0}}(0,t) O_{{\sm \Sc}} U_{{\sm 0}}(t,0), \nonumber \\
\hat \rho_{{\sm \In}}(t) &=& U_{{\sm 0}}(0,t) \hat\rho_{{\sm \Sc}}(t) U_{{\sm 0}}(t,0).
\end{eqnarray}
In the above equations the density matrices with hats are the total density operators associated with the total Hamiltonian $H$. An important formula we need to use for a perturbative expansion is the scattering operator, also known as the evolution operator in the interaction picture,
\begin{equation}
S(t,t') = \mathrm{T} \exp\left( -\frac{\iu}{\hbar} \int_{t'}^{t} V_{{\sm \In}}(t'') dt''\right),\quad t>t',
\end{equation}
where $V_{{\sm \In}}(t)$ is the coupling in the interaction picture. Using the definition of the scattering operator one easily obtains
\begin{eqnarray}
S(t,t') &=& U_{{\sm 0}}(0,t) U(t,t') U_{{\sm 0}}(t',0), \nonumber\\
\hat\rho_{{\sm \In}}(t) &=& S(t,t') \hat\rho_{{\sm \In}}(t') S(t',t). 
\end{eqnarray}

To demonstrate the generality of our theory in this paper we will consider three models: (I) - harmonic oscillator model with one degree of freedom in the system connected to harmonic baths.  For this model our Hamiltonian for the system is given by
\begin{equation}
\label{eq:HOC}
H_{{\sm S}}^{\mathrm{{\sm (I)}}}  = \frac{1}{2}p^2 + \frac{1}{2}\wz^2 u^2.
\end{equation}
Above we have set the mass to unity, or more precisely transformed the coordinate such that $u = x\sqrt{m}$, where $x$ is the usual displacement with dimension of length. The momentum $p$ above is conjugate to the coordinate $u$. $\wz$ is oscillator's angular frequency. The coupling between the system and the baths is assumed to be linear of the form
\begin{equation}
\label{eq:HOV}
 V^{\mathrm{{\sm (I)}}}  = u \sum_{j,\alpha=L,R} g_{\alpha,j}\, Q_{\alpha,j}, 
\end{equation} 
where $g_{\alpha,j}$ determines the strength of the system-bath coupling and $Q_{\alpha, j}$ is the position operator for the $j$-th oscillator of the harmonic baths. The baths are modeled as a collection of uncoupled harmonic oscillators with Hamiltonian 
\begin{equation}
\label{eq:HOB}
H_{\alpha}^{\mathrm{{\sm (I)}}} = \sum_{j} \left(\frac{1}{2}P_{\alpha,j}^2 + \frac{1}{2}\omega_{\alpha,j}^2\, Q_{\alpha,j}^2\right),
\end{equation}
where $\alpha = L, R$ corresponds to the left or right bath and $P_{\alpha,j}$ ($Q_{\alpha,j}$) is the momentum (conjugate position) for the $j$-th oscillator. The properties of the baths will be determined by the spectral density $J_{\alpha}(\omega)$, which will be required for concrete calculations later.

Model (II) will be the electron (spinless fermion) analogue of the harmonic oscillator model, also commonly known as the quantum dot (QD) model. In this case the system Hamiltonian is given by
\begin{equation}
\label{eq:QDC}
H_{{\sm S}}^{\mathrm{{\sm (II)}}}  = E_{{\sm 0}} d^\dagger d,
\end{equation} 
and the coupling 
\begin{equation}
\label{eq:QDV}
V^{\mathrm{{\sm (II)}}}  = \sum_{j,\alpha=L,R} g_{\alpha,j}\, c_{\alpha,j}^\dagger\, d + \mathrm{h.c.},
\end{equation}
where $d$ ($d^\dagger$) is the fermionic annihilation (creation) operator of the QD, and similarly $c$ ($c^{\dagger}$) is for the fermionic baths. Above, h.c. stands for Hermitian conjugate. The baths consist of non-interacting electrons and their Hamiltonian is given by
\begin{equation}
\label{eq:QDL}
H_{\alpha}^{\mathrm{{\sm (II)}}}  = \sum_{j} \varepsilon_{\alpha,j}\, c_{\alpha,j}^\dagger\, c_{\alpha,j}.
\end{equation}
In case of the QD model the total number of electrons in the baths is given by the number operator $N_{\alpha} = \sum_{j} c_{\alpha,j}^{\dagger}\,c_{\alpha,j}$. The properties of the baths will be determined using a spectral density $\Gamma_{\alpha}(E)$, which shall be fixed later.   

Both models (I) and (II) are linear (in terms of their equations of motion) and thus exactly solvable. They serve as a rigorous check, but the method outlined in this work does not exhibit any added advantage over the more powerful exact techniques, like the NEGF approach. Hence to demonstrate the wide applicability of our approach we take a nonlinear spin-boson model which does not admit an exact solution. The Hamiltonian of the system is then given by
\begin{equation}
\label{eq:SBC}
H_{{\sm S}}^{\mathrm{{\sm (III)}}}  = \frac{E}{2}\sigma_z + \frac{\Delta}{2} \sigma_x,
\end{equation}
where $\sigma_x$ ($\sigma_z$) is the $x$- ($z$-) component of the Pauli spin-1/2 matrices. The coupling to the baths will be assumed through the $\sigma_z$ operator of the system and is given by
\begin{equation}
\label{eq:SBV}
V^{\mathrm{{\sm (III)}}} =\frac{\sigma_{z}}{2}\sum_{j,\alpha=L,R} g_{\alpha,j} \,Q_{\alpha,j}.
\end{equation} 
The baths are considered to be a set of harmonic oscillators and take the same form as Eq.~(\ref{eq:HOB}).

\section{Dyson expansion}
\label{sec:3}
In this section we focus on the reduced density matrix defined via the Hubbard operator $X^{T}$. The Hubbard operator $X$ throughout this work will be defined as a matrix with elements $X_{mn} = X_{nm}^{T} = |m\rangle \langle n|$, where $|m\rangle$ is a ket in the eigenbasis of $H_{{\sm S}}$. Thus, working in the interaction picture with respect to $\h$ we have
\begin{eqnarray}
\label{eq:Xht}
\rho(t) = \overline{X^{T}_{{\sm \He}}(t)} &=& {\rm Tr}\bigl[ \rhI \rho_{{\sm L}} \rho_{{\sm R}} S(t_{{\sm 0}}, t) X^{T}_{{\sm \In}}(t) S(t,t_{{\sm 0}}) \bigr]  \\
&=& {\rm Tr}\left[ \rhI \rho_{{\sm L}} \rho_{{\sm R}} \mathrm{T_c} \left\{  X^{T}_{{\sm \In}}(t) e^{\lambda \int_c V_{{\sm \In}}(\tau)d\tau} \right\} \right], \nonumber
\end{eqnarray}
where $X^{T}_{{\sm \In}}(t) = e^{(\iu /\hbar) \, \h t} X^{T}_{{\sm \Sc}} e^{-(\iu /\hbar) \, \h t}$ is the Hubbard operator in the interaction picture. Above $\rhI$ is the density matrix of the system at initial time $t_{{\sm 0}}$ in the interaction picture and $\rho_{{\sm L,R}}$ are the bath density matrices, which are assumed to be of the canonical form. Assuming $\h$ to be time-independent, we can combine the two pieces of the $S$ operator [$S(t_{{\sm 0}}, t)$ and $S(t,t_{{\sm 0}})$] and consider a contour-time $\tau$ as demonstrated in the second line of Eq.~(\ref{eq:Xht}). The contour $C$ runs from $t_{{\sm 0}}$ to the time of interest $t$ and back to $t_{{\sm 0}}$, and $\mathrm{T_c}$ is the time-ordering super-operator on the contour. The adiabatic switch-on parameter $e^{\epsilon t}$ has been implicitly included in $V_{{\sm \In}}(t)$. The parameter $\lambda =(-\iu /\hbar)$ serves as a formal small expansion parameter in the Dyson series. Therefore, an expansion in $\lambda$ is equivalent to an expansion in the strength of the system-bath coupling $V$. For notational simplicity we will drop the subscripts $\Sc$, $\He$, and $\In$ representing the three pictures, and from the form of the operators it will be clear which picture they belong to.

In this work we will focus on the steady-state averages, i.e., averages at $t = 0$, and hence the steady-state average for the reduced density matrix $\rho$ can be computed as $\rho = \overline{X^{T}_{{\sm \He}}(0)}$. Performing the power series expansion for the exponential, and assuming that tracing over the baths for an odd power of the system-bath coupling operator gives zero, we obtain
\begin{equation}
\label{eq:dyson-expan-rho}
\rho = \langle X^T\rangle + \frac{\lambda^2}{2!}
\langle X^T V^2 \rangle + \frac{\lambda^4}{4!}\langle X^T V^4\rangle + \mathcal{O}(\lambda^6),
\end{equation}
where we have introduced a short-hand notation for the angle brackets $\langle \cdots \rangle$ $\equiv {\rm Tr}\bigl[ \rhI \rho_{{\sm L}} \rho_{{\sm R}} \mathrm{T_c} \int_c d\tau \cdots \bigr]$ and thus a generic term is given by 
\begin{widetext}
\begin{equation}
\label{eq:XVn}
\langle X^{T} V^n \rangle = {\rm Tr}\left[ \hat \rho(t_{{\sm 0}}) \int_c\!\!\! d\tau_1 \cdot d\tau_n \mathrm{T_c}  \Bigl\{X^{T}(0) V(\tau_1) \cdots V(\tau_n)\Bigr\} \right]
= {\rm Tr}\left[ \int dt_{1}\cdot dt_{n} \mathrm{T} \Bigl\{ X^{T}(0) [V(t_n), \cdots, V(t_1), \hat \rho(t_{{\sm 0}})] \Bigr\} \right],
\end{equation}
\end{widetext}
where all operators under the trace are in the interaction picture and $\hat{\rho}(t_{{\sm 0}}) = \rhI \rho_{{\sm L}} \rho_{{\sm R}}$. Above for the real-time integrals we have used the right normed convention for the nested commutators to avoid excess brackets, e.g., in case of $n = 3$ we get $[V(t_3),V(t_2),V(t_1),\hat\rho(t_{{\sm 0}})] \equiv [V(t_3),[V(t_2),[V(t_1),\hat{\rho}(t_{{\sm 0}})]]]$. Clearly, the number of contour integrals is the same as the power $n$ of $V^n = V(\tau_1) V(\tau_2) \cdots V(\tau_n)$. It is worth noting that in the equation above the parameters $\tau_i$ run on the contour with contour ordering among the operators $X^{T}(0)$ and $V(\tau_i)$, while the real time $t_i$ varies from initial time $t_{{\sm 0}}$ to final time $t=0$ and the time-arguments are time-ordered.

We now discuss one of our crucial observations that each term of the truncated-Dyson series (except the lowest) is infinite in the steady-state limit. The nature of this divergence is similar to a Taylor expansion of $e^{-t} = 1 -t + t^2/2 + \cdots$, when $t \to \infty$. To understand the origin of the divergences we prove in Append.\ref{append:A} a general result, valid for linear baths, given by
\begin{equation}
\label{eq:xv-eps}
\frac{1}{n} \langle X^{T} V^n \rangle_{\rho_d} = 
\frac{1}{\iu \fr + n \epsilon} \langle [X^{T}, V]V^{n-1} \rangle_{\rho_d}.
\end{equation}
Above the angle brackets have the same meaning as defined earlier, except that $[X^{T}, V]$ is taken at time $t=0$ and the subscript $\rho_d$ implies that we focus on the diagonal elements of the reduced density matrix of the system, $\rhI$. We have also taken the limit $t_{{\sm 0}} \to -\infty$ above and kept a finite adiabatic switch-on parameter $\epsilon$. The steady-state limit of the above equation can be easily obtained by taking the limit $\epsilon \to 0^+$. Above, $\fr$ is a super-operator and it has a special interpretation which can be correctly understood in the eigenbasis of $H_{{\sm S}}$. The operator $\fr$ contains information about the eigenenergy differences of $H_{{\sm S}}$ and takes a particular value of $\fr= (E_m-E_n)/\hbar$ ($E_m$ is the $m$-th eigenenergy of $H_{{\sm S}}$ with eigenvector $|m\rangle$) only when the Hubbard operator $X$ takes the form $X = |m\rangle\langle n|$ within the angle brackets. It is then clear that divergences appear at each order of the Dyson expansion [left-hand side of Eq.~(\ref{eq:xv-eps})], whenever the operator $\fr$ is equal to zero in the eigenbasis of $H_{{\sm S}}$ or strictly the terms diverge as $\mathcal{O}(1/\epsilon)$ with $\epsilon \to 0^+$. Hence, the Dyson series can be viewed as a power series in $1/\epsilon$ to arbitrary high powers. The proof of the above identity requires the time-translational invariance of the bath correlators. Thus, the physical origin of the divergence is because the system becomes time translationally invariant in the limit $t_{{\sm 0}} \to -\infty$ and $\epsilon \to 0^+$. Although throughout this section we have focused on the reduced density matrix $\rho$ the same conclusions can be drawn for any general operator $O$. Thus, for any observable the truncated-Dyson series would lead to divergences in the steady-state limit at each order of the expansion.
\section{Determining the initial reduced density matrix}
\label{sec:4}
In principle, the divergences in the Dyson series can be removed by summing the entire series. But in practice, truncations are un-avoidable and here we take the point of view that the divergences may be ``canceled'' term by term by a proper choice of the initial system density matrix $\rhI$. Any other generic, arbitrary, but finite choice of $\rhI$ leads to divergent terms in the series as shown in Sec.\ref{sec:3}. 

Thus to obtain the correct choice of $\rhI$, which removes the divergences at each order, we start from the Dyson expansion of the reduced density matrix Eq.~(\ref{eq:dyson-expan-rho}) and take the limit $t_{{\sm 0}} \to -\infty$ keeping a finite $\epsilon$. This physically implies that we are not working in the steady-state limit. Hence a unique one-to-one map exists between the reduced density matrix $\rho$, which is correct upto all orders and the initial reduced density matrix $\rhI$. This map can be inverted recursively, using Eq.~(\ref{eq:dyson-expan-rho}), as long as we use a finite $\epsilon$ and upto $6$th order it is given by
\begin{eqnarray}
\rhI &=& \rho - \frac{\lambda^2}{2} \langle X^T V^2\rangle_\rho 
-\frac{\lambda^4}{4!}\langle X^T V^4 \rangle_\rho \nonumber \\
&&+ \frac{\lambda^4}{(2!)^{2}} \bigl\langle \langle X^T V^2 \rangle_\rho X^T V^2 \bigr\rangle - \frac{\lambda^6}{6!} \langle X^T V^6\rangle_\rho  \nonumber \\
&&+ \frac{\lambda^6}{2!\,4!}\Biggl[ \bigl\langle \langle X^T V^2 \rangle_\rho X^T V^4 \bigr\rangle +  
 \bigl\langle \langle X^T V^4 \rangle_\rho X^T V^2 \bigr\rangle\Biggr] \nonumber \\ 
\label{eq-rho0}
&&- \frac{\lambda^6}{(2!)^3} \bigl\langle \bigl\langle \langle X^T V^2 \rangle_\rho X^T V^2 \bigr\rangle X^T V^2 \bigr\rangle  + \mathcal{O}(\lambda^8),  
\end{eqnarray}
where we have introduced two types of angle brackets. The brackets at the innermost level with a subscript $\rho$ are the same as before except that $\rhI$ is replaced by $\rho$. The outer slightly larger angle brackets mean a trace over the density matrix of the bath, $\rho_{{\sm L,R}}$, as well as over the system with an auxiliary matrix produced by $\langle \cdots \rangle$ inside it. In order to understand these double angle brackets better let us consider the term $\bigl\langle \langle X^T V^2 \rangle_\rho X^T V^2 \bigr\rangle$. Here we first evaluate the innermost angle bracket $\langle X^T V^2\rangle_{\rho} \equiv \tilde{\varrho}$, whose elements can be evaluated as $\tilde{\varrho}_{mn} = {\rm Tr}\Bigl[\rho \rho_{{\sm L}} \rho_{{\sm R}} |n\rangle \langle m|  \int_c d\tau_1 d\tau_2 \mathrm{T_c}\bigl\{ V(\tau_1) V(\tau_2)  \bigr\} \Bigr]$. Then the elements of the outermost bracket $\bigl\langle \langle X^T V^2 \rangle_\rho X^T V^2 \bigr\rangle_{mn} = 
{\rm Tr}\Bigl[\tilde{\varrho} \rho_{{\sm L}} \rho_{{\sm R}} |n\rangle \langle m|  \int_c\! d\tau_1 d\tau_2 \mathrm{T_c} \bigl\{V(\tau_1) V(\tau_2)\bigr\}  \Bigr]$, which is implicitly a linear function of $\rho$.

The above mapping is in the same spirit as that of van Kampen \cite{Kampen174, Kampen274}, who used a similar inversion scheme to obtain the time-local quantum master equation (TLQME) \cite{fulinskiphysica68, lugiatophysica70, shibatajstatphys77, shibatajpsj80}. Clearly, even though we have obtained an equation for $\rhI$, Eq.~(\ref{eq-rho0}), we have not solved the problem completely because we still need to know the \emph{exact} reduced density matrix, $\rho$. Interestingly, the TLQME gives an \emph{exact} differential equation for $\rho$ in the interaction picture \cite{laird91},
\begin{eqnarray}
\frac{d\rho}{d t} &=& \frac{d \Phi(t,t_{{\sm 0}})}{d t}\Phi(t,t_{{\sm 0}})^{-1}\rho,
\label{eq:tlqe}
\end{eqnarray}
where
\begin{eqnarray}
\Phi(t,t_{{\sm 0}}) &=& \mathrm{Tr}_{{\sm L,R}}\left(\mathrm{T}\left\{e^{\int_{t_{{\sm 0}}}^{t}dt'\, \mathcal{L}(t')}\right\}\rho_{{\sm L}}\rho_{{\sm R}}\right),
\end{eqnarray}
with $\mathcal{L}(t)$ being the Liouvillian super-operator in the full Hilbert space, namely $\partial \hat{\rho}/\partial t = \mathcal{L}(t)\hat{\rho}$. It is important to stress here that the initial condition for the total density matrix is factorized, i.e., $\hat{\rho}(t_{{\sm 0}}) = \rhI \otimes \rho_{{\sm L}} \otimes \rho_{{\sm R}}$, whereas Eq.~(\ref{eq-rho0}) represents the reduced density matrix of the system. We could in principle solve the above equation to obtain $\rho$ and hence  obtain $\rhI$. Practically this is a cumbersome task and no exact solution exists for general nonlinear systems. Hence for practical feasibility we will herein exploit the linearity of Eq.~(\ref{eq-rho0}). If we would like to evaluate the average of any observable $O$, as given by the truncated-Dyson expansion upto the $N$th order in $\lambda$, then due to the linearity of the truncated-Dyson expansion we would require $\rhI$ correct only upto the $N$th order. This fact translates in requiring the reduced density matrix $\rho$ correct upto $\mathcal{O}(\lambda^N)$, due to the linearity of Eq.~(\ref{eq-rho0}). Thus, clearly we do not require the \emph{exact} reduced density matrix $\rho$ but its Taylor series expansion correct upto order $N$. Recently it has been shown \cite{fleming11, thingna-jcp12} that in order to obtain the Taylor series expansion of the reduced density matrix $\rho$ correct upto $\mathcal{O}(\lambda^{N})$ one would require to solve the QME of $\mathcal{O}(\lambda^{N+2})$. This fact will be explored further in Sec.~\ref{sec:5} to evaluate currents upto fourth order in system-bath coupling strength. 

In Eq.~(\ref{eq:tlqe}) we have still not taken the mathematically correct steady-state limit of $\epsilon \to 0^+$. Despite this we have imposed the condition that $d\rho/dt = 0$, which would then ensure that $\rho$ is the correct steady-state solution. This physically implies that we search for a specific $\rhI$ which in long, but \emph{finite}, time leads to the steady state. This hypothesis of choosing a $\rhI$ based on Eqs.~(\ref{eq-rho0})-(\ref{eq:tlqe}) constitutes our first important result of this work. Our approach resolves another important issue that the steady-state averages obtained via the improved Dyson series expansion would no longer depend on \emph{arbitrary} initial conditions. It is important to note that keeping a finite $\epsilon$ at all the intermediate steps is crucial to remove the divergences and this will be shown using a concrete example of the current-operator in Sec.~\ref{sec:5}.

\section{Fourth-order Currents}
\label{sec:5}
In this section we take the specific example of the current-operator and evaluate currents upto $4$th order in system-bath coupling. We then explicitly demonstrate that our approach to fix the initial condition $\rhI$ correctly cancels the divergences at the $4$th order. 

The mathematical structure of the current is the same as that of the QME for $\rho$, see Eq.~(\ref{eq:4qme}) below, except that we need to replace the commutator $[X^T,V]$ by $\leftdot{V}$, where the left-sided dot on $V$ indicates that the time derivative is performed only with respect to the left bath and the factor of $-\iu/\hbar$ is omitted from the Heisenberg equation of motion. Then we can write both energy, $I_{{\sm L}}^{en} = -\overline{d H_{{\sm L}}/dt} = - \mathrm{Tr}\left[d H_{{\sm L}}/dt\,\hat{\rho}\right]$, and electronic, $I_{{\sm L}}^{el} = -\overline{d N_{{\sm L}}/dt} = - \mathrm{Tr}\left[d N_{{\sm L}}/dt\,\hat{\rho}\right]$ (in units of elementary charge $e$), currents into one unified notation $\mathcal{I}_{{\sm L}} = \lambda \overline{\leftdot{V}} = \lambda\mathrm{Tr}[\leftdot{V}\hat{\rho}]$, where $\leftdot{V} =u\sum_{j} g_{{\sm L},j}\,\iu\hbar P_{{\sm L},j}$ for energy current in models (I), and (III) and $\leftdot{V} = -\sum_{j} g_{{\sm L},j}\, c_{{\sm L},j}^\dagger\, d - {\rm h.c.}$ for electronic current in model (II). Thus the current $\mathcal{I}_{{\sm L}}$ upto $4$th order in $\lambda$ is given by
\begin{eqnarray}
\mathcal{I}_{{\sm L}} &=& \lambda \mathrm{Tr}\left[\rho(t_{{\sm 0}}) \rho_{{\sm L}} \rho_{{\sm R}} \mathrm{T_c} \leftdot{V} e^{\lambda \int_c V(\tau) d\tau} \right] \\
&\approx & \lambda^2 \langle \leftdot{V}V \rangle + 
\frac{\lambda^4}{3!} \langle \leftdot{V}V^3 \rangle \nonumber \\
\label{eq:4curr}
&\approx & \lambda^2 \langle \leftdot{V}V \rangle_{\rho} \! + \!
\frac{\lambda^4}{3!} \langle \leftdot{V}V^3 \rangle_{\rho}
\!-\!\frac{\lambda^4}{2!} \bigl\langle \langle X^TV^2\rangle_{\rho} \leftdot{V} V \bigr\rangle .
\end{eqnarray}
The first line expresses the problem in the interaction picture and the second line is the Dyson expansion.  Here since $\leftdot{V}$ contains one bath operator we need to keep only the odd powers in $V$. The last line is the improved Dyson series approach, due to Eq.~(\ref{eq-rho0}), where $\rhI$ is written in terms of $\rho$. In Eq.~(\ref{eq:4curr}) a divergence appearing in the second term gets canceled explicitly by the third term. The current expression, Eq.~(\ref{eq:4curr}), and the QME, Eq.~(\ref{eq:4qme}) below, show no divergences because they are in the form of \emph{ordered} cumulants \cite{kubojpsj62, mukamelpra88}, which remain convergent as shown by van Kampen \cite{Kampen174, Kampen274}. Also since the reduced density matrix at time $t=0$ is well behaved, we can now take the limit $\epsilon \to 0^+$ in the last line of the current expression, and obtain a finite result. This equation is our central result which generalizes the commonly used second-order result~\cite{wupre09, juzar-prb12} (the first term). Since the current is expressed in terms of the exact reduced density matrix $\rho$, we must solve the $4$th order QME in order to obtain the current. 

Since we require current accurate upto $4$th order in $\lambda$, we can see from Eq.~(\ref{eq:4curr}) that we require the reduced density matrix correct upto $2$nd order. This can be achieved using Eq.~(\ref{eq:tlqe}) to obtain the $4$th order TLQME as
\begin{eqnarray}
\frac{d\rho}{dt} &=&  -\frac{\iu}{\hbar} \left[H_{{\sm S}},\rho\right] + 
\lambda^2 \langle [X^T,V]V \rangle_{\rho} + 
\frac{\lambda^4}{3!} \langle [X^T,V]V^3 \rangle_{\rho} \nonumber \\
&&-\frac{\lambda^4}{2!} \bigl\langle \langle X^TV^2\rangle_{\rho} [X^T,V] V \bigr\rangle + \mathcal{O}(\lambda^6) = 0.
\label{eq:4qme}
\end{eqnarray}
Above, the time argument for $[X^T,V]$ is at $t=0$, while all the other $V$'s have dummy contour-time argument $\tau_i$ which need to be integrated out. The $\rho$ dependence is in the angle brackets, $\langle \cdots \rangle_{\rho} = {\rm Tr}[ \rho \rho_{{\sm L}} \rho_{{\sm R}} \mathrm{T_c} \int_c d\tau \cdots]$. After performing the complete trace and integrating over the contour, we obtain explicitly the equation for $\rho$. If we truncate the above equation upto $2$nd order we get the standard Bloch-Redfield QME \cite{redfield57}. 

The steady-state solution to the $4$th order QME is a highly non-trivial problem and till date has been achieved only for the case of equilibrium spin-boson model \cite{laird91, sjang02, Divincenzoprb05} and the nonequilibrium interacting quantum dot model \cite{timmprb08, kollerprb10}. In case of the equilibrium spin boson model Laird \emph{et~al.} \cite{laird91} calculate the $4$th order relaxation tensor under the rotating wave approximation since they are interested in comparison with the Bloch equations. Jang \emph{et~al} \cite{sjang02} make a high temperature approximation for the ohmic baths to obtain analytical results, whereas Di Vincenzo and Loss \cite{Divincenzoprb05} study the problem in the opposite regime when temperature of the bath $T = 0$. For the nonequilibrium problem Koller \emph{et~al.} \cite{kollerprb10} obtain an expression in the time non-local form for the $4$th order electronic current which has a different diagrammatic structure as compared to the time local form expressed in this work \cite{sjang02, timmprb08}. Additionally, they do not solve the reduced density matrix order-by-order (as described below), which could drastically reduce the computational complexity thus allowing us to solve the bosonic problem with 100's of levels efficiently.

In this work, instead of attempting to solve Eq.~(\ref{eq:4qme}) by brute force we will outline an approach to solve it order-by-order \cite{thingna-jcp12} in the expansion parameter $\lambda$. To this end, we expand the reduce density matrix as
\begin{equation}
\rho = \rho^{(0)} +\lambda^2 \rho^{(2)} + \lambda^4 \rho^{(4)} + \cdots.
\end{equation}
Substituting the above expansion into the QME, and comparing powers of $\lambda$, we obtain for the $0$th, $2$nd and $4$th power in $\lambda$:
\begin{eqnarray}
\label{eq:drho1}
&-&\frac{\iu}{\hbar} [H_{{\sm S}},\rho^{(0)}] = 0, \\
\label{eq:drho2}
&-&\frac{\iu}{\hbar} [H_{{\sm S}},\rho^{(2)}] +  \langle [X^T,V]V \rangle_{\rho^{(0)}} = 0,\\
\label{eq:drho4}
&-&\frac{\iu}{\hbar} [H_{{\sm S}},\rho^{(4)}] + 
 \langle [X^T,V]V \rangle_{\rho^{(2)}} + 
\frac{1}{3!}\langle [X^T,V]V^3 \rangle_{\rho^{(0)}}\nonumber \\
&&\qquad -\frac{1}{2!} \bigl\langle \langle X^TV^2\rangle_{\rho^{(0)}} [X^T,V] V \bigr\rangle  = 0.
\end{eqnarray}
Above we have imposed the steady state condition $d\rho/dt = 0$. We now split the $\rho^{(i)}$ and $X^T$ into diagonal (subscript $d$) and off-diagonal (subscript $f$) parts in the eigenbasis of $H_{{\sm S}}$ as $\rho^{(i)} = \rho^{(i)}_d + \rho^{(i)}_f$, $i=0, 2, 4,\cdots$, and $X^T = X^T_d + X^T_f$. The Eqs.~(\ref{eq:drho1})--(\ref{eq:drho4}) are matrix equations which can be solved using a leap-frog method from diagonal to off-diagonal then to diagonal, from the lowest to the highest order. Setting $X^T$ to diagonal (off-diagonal) generates the diagonal (off-diagonal) part of the component equations. Then the off-diagonals can be solved trivially if the results in the lower order have already been obtained as
\begin{eqnarray}
\label{eq:qme2}
\rho^{(0)}_f &=& 0, \\
\label{eq:qme4}
-\frac{\iu}{\hbar} [H_{{\sm S}},\rho_{f}^{(2)}] &=& \langle [X^T_f,V]V \rangle_{\rho^{(0)}}.
\end{eqnarray}
The diagonal parts of the equations are linear equations given by
\begin{eqnarray}
\label{eq:qme1}
\langle [X^T_d,V]V \rangle_{\rho^{(0)}_d} &=&0, \\
\label{eq:qme3}
\langle [X^T_d,V]V \rangle_{\rho^{(2)}_d} &=& -\langle [X^T_d,V]V \rangle_{\rho^{(2)}_f}
-\frac{1}{3!} \langle [X^T_d,V]V^3 \rangle_{\rho^{(0)}} \nonumber \\
&&+ \frac{1}{2!} \bigl\langle \langle X^T V^2\rangle_{\rho^{(0)}} [X^T_d,V] V \bigr\rangle. 
\end{eqnarray}

In addition we require the normalization condition, ${\rm Tr}(\rho)=1$, to fix a unique solution. Thus, in order to evaluate the current correct upto $4$th order in $\lambda$ we first obtain the reduced density matrix correct upto $2$nd order using Eqs.~(\ref{eq:qme2})--(\ref{eq:qme3}). Then using $\rho = \rho^{(0)} + \lambda^2 \rho^{(2)}$ in Eq.~(\ref{eq:4curr}) we obtain the steady-state current correct upto the $4$th order. Alternatively, the $2$nd order reduced density matrix can be obtained via the modified Redfield solution approach \cite{Thingna13}, which reproduces the correct reduced density matrix for linear systems. Here we will not resort to this approach and instead solve the $4$th order quantum master equation as described above. 

\begin{figure}
 \centering
\includegraphics[width=\columnwidth]{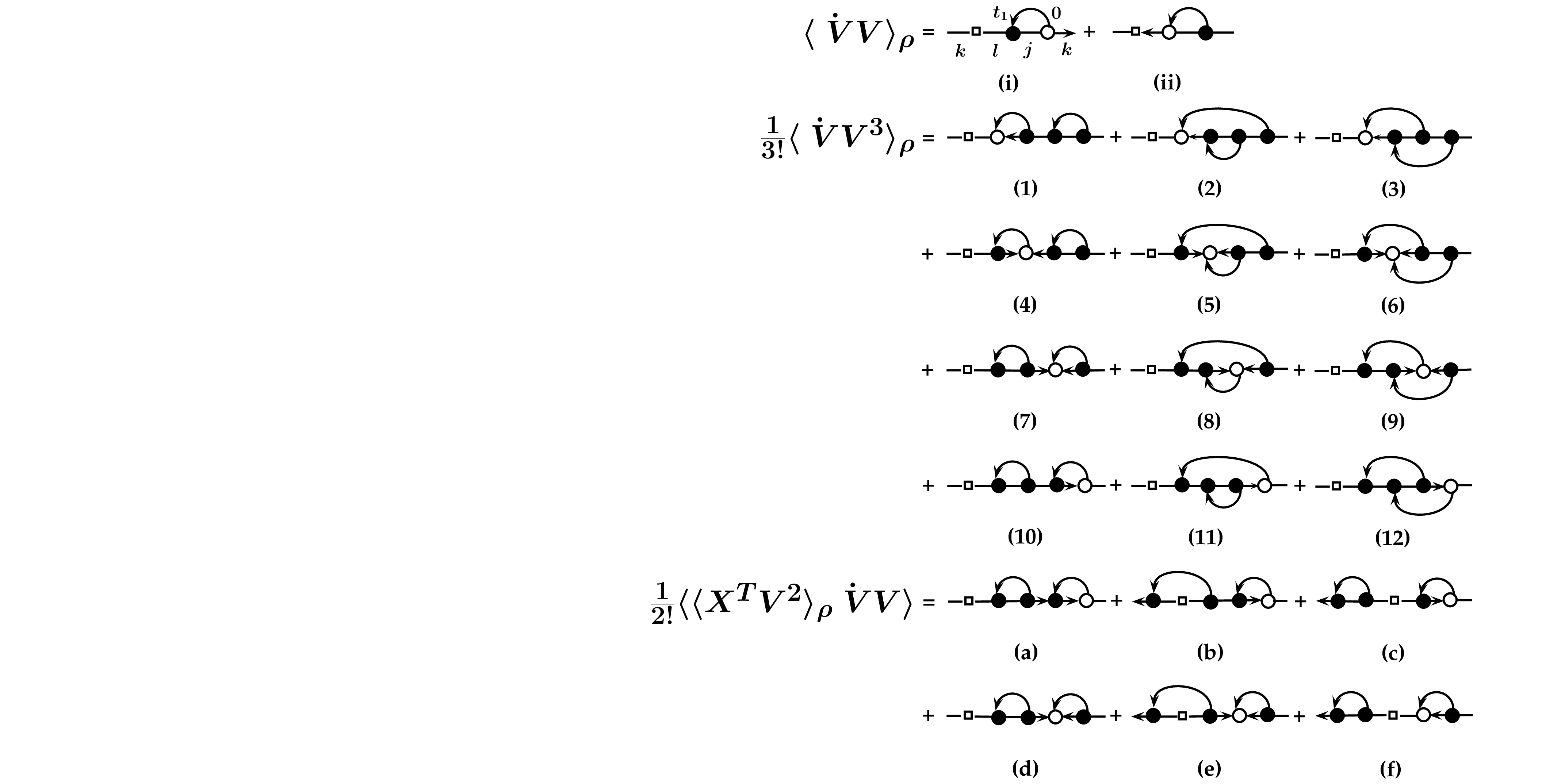}
\caption{\label{fig:diagram}Diagrams representing the terms for the current. The open dots represent $\leftdot{V}$ for currents (or $[X^T,V]$ for the quantum master equation). The diagrams (1), (4), (5), (7), (8), (10) and (a)-(f) have divergent terms of the form $\propto 1/\epsilon$.  Note that (c) cancels (4), and (d) cancels (7) exactly. The Feynman rules are discussed in the text.}
\end{figure}

\subsection*{Diagrammatics}
\label{sec:5A}
In order to organize the calculations of various expansion terms appearing in the QME or the current, it is useful to use diagrams (for similar diagrammatic rules see ref.~[\onlinecite{kollerprb10}]) to represent the algebraic structure of the terms as shown in Fig.~\ref{fig:diagram}. In this section to create the diagrams we will consider the case when both baths are connected to the same degree of freedom in the system and the coupling operator $V$ is given by a generic form $V = \sum_{\alpha,\beta} S^{\alpha,\beta} \otimes B^{\alpha,\beta}$, where $\alpha$ takes a sum over the number of baths, i.e., $\alpha = L,R$ and $\beta$ is a general sum over the number of system and bath operators involved. The operator $S^{\alpha,\beta}$ resides in the system Hilbert space and $B^{\alpha,\beta}$ belongs to the Hilbert space of the baths. The operator $\leftdot{V} = \sum_{\beta} S^{{\sm L},\beta} \otimes \dot{B}^{{\sm L},\beta}$, where $\dot{B}^{{\sm L},\beta}$ is an operator in the left-bath Hilbert space and it depends on the form of the current operator.

In order to create the diagrams we follow two basic steps. The first step is to unravel the contour time into normal time with time or anti-time order. This will be represented by a horizontal arrow with dots. Each dot has a particular time variable $t_i$ and the operator $V$ associated with it. An open dot denotes a particular time of $0$ and associated with it is the operator $\leftdot{V}$ (in case of the density matrix the operator $\leftdot{V}$ is to be replaced by $[X^T, V]$). The times on the left pointing arrows are ordered (when read from right to left), and on right pointing arrows are anti-time ordered. In any event, the arrows are drawn to point from $-\infty$ to $0$. Any horizontal link between the dots has a dummy system state label, such as those indicated in diagram (i). A square box represents the density matrix $\rho$ sitting at a time of $-\infty$. The state labels are summed so as to imply a matrix multiplication and trace. Hence, for example, if the same system operator connects to the left and right baths, i.e., $S^{{\sm L},\beta} \equiv S^{{\sm R},\beta} = S$, then bath operators take the form $B^{{\sm L},\beta} = B^{{\sm L}}$ and $B^{{\sm R},\beta} = B^{{\sm R}}$. Thus, the diagram (i) in Fig.~\ref{fig:diagram} represents (for the system part):
\begin{equation}
{\rm Tr} \bigl[\rho S(t_1) S(0)\bigr] = \sum_{k,l,j} \rho_{kl} S_{lj}(t_1) S_{jk}(0).
\end{equation}
The second step is to apply Wick's theorem to connect the dots in all possible ways and ensuring that the baths are uncorrelated, i.e., $\mathrm{Tr}_{{\sm L,R}}\left(B^{{\sm L}}(t)B^{{\sm R}}(0)\rho_{{\sm L}}\rho_{{\sm R}}\right) = \mathrm{Tr}_{{\sm L,R}}\left(\dot{B}^{{\sm L}}(t)B^{{\sm R}}(0)\rho_{{\sm L}}\rho_{{\sm R}}\right) = 0$. Finally, all the dummy time variables need to be integrated out.

Using the above rules the diagram (i) for the current reads
\begin{equation}
\mathrm{dia.~(i)} = \sum_{k,l,j}\rho_{kl} S_{lj} S_{jk}  \int_{-\infty}^{0} dt_{1} e^{\iu \fr_{lj} t_{1} + \epsilon t_{1}} \chi_{{\sm L}}(t_{1}),
\end{equation}
where $\chi_{{\sm L}}(t_{1}) = -\mathrm{Tr}_{{\sm L}}\left(\dot{B}^{{\sm L}}(t_{1})B^{{\sm L}}(0)\rho_{{\sm L}}\right)$. Above we have used $S_{nm}(t_{1}) = e^{\iu\fr_{nm}t_{1}}S_{nm}$, where $\fr_{nm} = \left(E_{n}-E_{m}\right)/\hbar$, and $\mathrm{Tr}_{{\sm L,R}}\left(\dot{B}^{{\sm L}}(t_{1})B^{{\sm R}}(0)\rho_{{\sm L}}\rho_{{\sm R}}\right) = 0$. As another example let us consider diagram (3) which is given by
\begin{eqnarray} 
 && -\sum_{\substack{k,l,p,q,j \\ \alpha={\sm L,R}}} \rho_{kl} S_{lp}S_{pq}S_{qj}S_{jk} \int_{-\infty}^{0}dt_{1}
\int_{-\infty}^{t_{1}}dt_{2} \int_{-\infty}^{t_{2}}dt_{3}  \nonumber \\
&& e^{\iu (\fr_{pq}t_{1}  +\fr_{qj}t_{2} + \fr_{jk}t_{3}) + \epsilon(t_{1} + t_{2} +t_{3})} \chi_{{\sm L}}(-t_{2}) C_{{\alpha}}(t_{1}-t_{3}), \nonumber  
\end{eqnarray}
where $C_{\alpha}(t) = \mathrm{Tr}_{\alpha}\left(B^{\alpha}(t)B^{\alpha}(0)\rho_{\alpha}\right)$ is the correlator of the baths. The divergent terms can be easily recognized from the diagrams in Fig.~\ref{fig:diagram} by carefully observing their structure. A term will diverge if the corresponding diagram contains a time-translationally invariant part, e.g., in case of diagram (4) we can move the two solid dots to $-\infty$ without disturbing the other dots. The diagrams which are entangled with the open dot (which is at a fixed time $0$) are always finite, e.g., diagram (3). The infinities in diagrams (1) -- (12) get precisely canceled by the subtracting terms from diagrams (a) -- (f).  

For the QD model [model (II)], since each lead has two distinct operators connecting to the system, we have two distinct bath correlators. Therefore, the number of diagrams are doubled for $2$nd order terms and quadrupled for $4$th order. We can take the convention that left pointing arrows on the diagrams are associated with one of the bath-correlator $C^<$ and right ones with $C^>$ [See Eqs.~(\ref{eq:QDCg}) and (\ref{eq:QDCl}) for definitions of these correlators]. The arrows on the bath correlators start from a system creation operator $d^\dagger$ and end at an annihilation operator $d$.  Since for a one-degree system $d^2 = (d^\dagger)^2 = 0$, a head next to a head or a tail next to a tail of the arrows is not allowed, which reduces the number of nonzero diagrams. 
\section{Archetypal examples}
\label{sec:6}
In this section we deal with some of the most common examples of transport problems as given by the three models described in Sec.\ref{sec:2}. Our goal here would be to compare with the exact NEGF formalism \cite{wang08review, wang13review} for the exactly solvable harmonic oscillator system connected to harmonic baths (model I) and a QD connected to fermionic baths (model II). In the end we will tackle the nonlinear spin-boson model (model III). Throughout this section we will employ a combination of analytical and numerical techniques to obtain the currents for these systems upto $4$th order in system-bath coupling. 

\subsection{Harmonic oscillator model}
\label{sec:6A}
We start with the exactly solvable harmonic oscillator model whose Hamiltonian is described by Eqs.~(\ref{eq:HOC})--(\ref{eq:HOB}). Since there is no flow of particles in the system the energy current is synonymous to the heat current and hence the heat current operator is given by
\begin{equation}
\hat{I}_{{\sm L}}^{en} = -\frac{d H_{{\sm L}}}{dt} = -\frac{i}{\hbar}
[H,H_{{\sm L}}] = \lambda \leftdot{V},
\end{equation}
where $H$ is the total Hamiltonian and $H_{{\sm L}}$ is the Hamiltonian of the left-bath given by Eq.~(\ref{eq:HOB}). The operator $\leftdot{V}$ can be expressed as 
\begin{equation}
\leftdot{V} = u\,\iu\hbar\sum_{j} g_{{\sm L},j}\,P_{{\sm L},j}.
\end{equation}

The baths are completely described by their spectral density which in terms of the bath parameters can be described as
\begin{eqnarray}
J_{\alpha}(\omega)&=&\frac{\pi}{2}\sum_{j}\frac{|g_{\alpha, j}|^{2}}{\omega_{\alpha, j}}\delta(\omega-\omega_{\alpha, j}).
\end{eqnarray}
In order to calculate the currents and the reduced density matrix we require two different types of bath correlators which can be expressed in terms of the spectral density as
\begin{eqnarray}
\label{eq:CoorC}
C_{\alpha}(t) &=& \sum_{j}|g_{\alpha,j}|^{2}\mathrm{Tr}_{\alpha}\left(\rho_{\alpha}Q_{\alpha,j}(t)Q_{\alpha,j}(0)\right)\nonumber \\
&=& \frac{\hbar}{\pi}\int_{-\infty}^{\infty}d\omega J_{\alpha}(\omega)n_{\alpha}(\omega)e^{\iu\omega t},\\
\label{eq:CoorChi}
\chi_{{\sm L}}(t) &=& -\sum_{j}|g_{{\sm L},j}|^{2}\mathrm{Tr}_{{\sm L}}\left(\rho_{{\sm L}}P_{{\sm L},j}(t)Q_{{\sm L},j}(0)\right)\nonumber \\
&=& -\frac{\iu \hbar}{\pi}\int_{-\infty}^{\infty}d\omega\,\omega J_{\alpha}(\omega)n_{\alpha}(\omega)e^{\iu \omega t},
\end{eqnarray}
where $n_{\alpha}(\omega) = [e^{\beta_{\alpha}\hbar\omega}-1]^{-1}$ is the Bose-Einstein distribution function containing the temperature information of the baths. The forms given in terms of the spectral density assume $J_{\alpha}(-\omega)=-J_{\alpha}(\omega)$. The correlators above are related via a time derivative, i.e., $\chi_{{\sm L}}(t) = -dC_{{\sm L}}(t)/dt$, and since we are interested in the heat current flowing out of the left bath $\chi_{{\sm R}}$ does not enter the calculations. The Fourier-Laplace transforms of these bath correlators would be of central interest and are given by
\begin{eqnarray}
\label{eq:W}
W_{pq}^{n} &=&\sum_{\alpha={\sm L,R}}\int_{-\infty}^{0}dt\,e^{\iu\fr_{pq}t}C_{\alpha}(t)e^{n\epsilon t}, \\ 
\label{eq:Wdot}
\dot{W}_{pq}^{n} &=& \int_{-\infty}^{0}dt\,e^{\iu\fr_{pq}t}\chi_{{\sm L}}(t)e^{n\epsilon t},\\
\label{eq:WWt1}
\dot{W}_{pq}^{n}W_{kl}^{m}[t_{{\sm 1}}] &=& \sum_{\alpha = {\sm L,R}} \int_{-\infty}^{0}dt_{{\sm 1}} e^{(\iu\fr_{pq}+n\epsilon)t_{{\sm 1}}}\chi_{{\sm L}}(t_{{\sm 1}}) \nonumber \\
&&\times \int_{t_{{\sm 1}}}^{0}dt_{{\sm 2}} e^{(\iu\fr_{kl}+m\epsilon)t_{{\sm 2}}}C_{\alpha}(t_{{\sm 2}}),
\end{eqnarray}
where $\fr_{pq} = (p-q)\wz$ are the energy differences of the harmonic oscillator.

Thus, for the harmonic oscillator case the $2$nd order heat current in the eigenbasis of $H_{{\sm S}}$  can be written as
\begin{eqnarray}
\label{eq:2hcurr}
\lambda^2 \langle \leftdot{V} V \rangle_{\rho} & = &\lambda \sum_{k,l,j}\rho_{kl}\,u_{lj}\,u_{jk}\left((\dot{W}_{kj}^{1})^{*}-\dot{W}^{1}_{lj}\right),
\end{eqnarray}
where $\rho_{ij}$ are elements of the exact reduced density matrix and the elements $u_{ij} = \langle i| u | j \rangle$. As described in Sec.\ref{sec:5} we will substitute the exact reduced density matrix with $\rho = \rho^{(0)}+\lambda^2\rho^{(2)}$ to obtain heat current accurate upto $4$th order in system-bath coupling. If one uses $\rho = \rho^{(0)}$ in Eq.~(\ref{eq:2hcurr}) then we recover the weak coupling result obtained previously \cite{segalprb06, juzar-prb12} using QMEs. 

In order to evaluate the $4$th order terms we use the diagrams illustrated in Fig.~\ref{fig:diagram}. We first notice that the diagrams (4) and (c), and diagrams (7) and (d) cancel each other exactly. The divergence in diagram (1) is canceled by diagram (f) and we group this as one term given by
\begin{equation}
\label{eq:dia1f}
\mathrm{dia.~(1-f)} = \sum_{\substack{p,q,j \\ k,l}}\rho_{kl}\,u_{pq}\,u_{qj}\,u_{lp}\,u_{jk} D^{pqj}_{lk}(1-f),
\end{equation}
where
\begin{eqnarray}
\label{eq:diaD1f}
D^{pqj}_{lk}(1-f)&=&\sum_{\alpha = {{\sm L,R}}} \int_{-\infty}^{0}dt_{{\sm 1}}\int_{t_{{\sm 1}}}^{0}dt_{{\sm 2}}\int_{-\infty}^{t_{{\sm 2}}}dt_{{\sm 3}} \,e^{\epsilon(t_{{\sm 1}}+t_{{\sm 2}}+t_{{\sm 3}})}\nonumber \\
 &&e^{\iu(\fr_{pq}t_{{\sm 1}}+\fr_{qj}t_{{\sm 2}}+\fr_{jk}t_{{\sm 3}})} \chi^{*}_{{\sm L}}(t_{{\sm 1}})C_{\alpha}(t_{{\sm 2}}-t_{{\sm 3}}).
\end{eqnarray}
Above $\chi^{*}_{{\sm L}}(\tau)$ denotes the complex conjugate of $\chi_{{\sm L}}(\tau)$. The triple integral can be simplified by making a change of variables $t_{{\sm 2}}^{\prime} = t_{{\sm 2}}-t_{{\sm 3}}$ and $t_{{\sm 3}}^{\prime} = t_{{\sm 2}}+t_{{\sm 3}}$ and then carrying out the $t_{{\sm 3}}^{\prime}$ integral analytically to obtain
\begin{equation}
D^{pqj}_{lk}(1-f) = \frac{1}{\iu\fr_{qk}+2\epsilon}\left(\dot{W}^{3}_{kp}W^{1}_{kj}-\dot{W}^{1}_{qp}W^{1}_{kj}\right)^{*}.
\end{equation}

Similarly diagrams $\delta =$ (2), (3), (8-b), and (9) can be expressed as
\begin{eqnarray}
\label{eq:dia2}
\mathrm{dia.~(\delta)} &=& \sum_{\substack{p,q,j \\ k,l}}\rho_{kl}\,u_{pq}\,u_{qj}\,u_{lp}\,u_{jk} D^{pqj}_{lk}(\delta),
\end{eqnarray}
where the $D$-functions are given by:
\begin{eqnarray}
\label{eq:diaD2}
D^{pqj}_{lk}(2)&=& \left[\iu\fr_{pj}+2\epsilon\right]^{-1} \nonumber \\
&&\times\left(\dot{W}^{1}_{kj}W^{1}_{jq}[t_{{\sm 1}}]-\dot{W}^{3}_{kp}W^{-1}_{pq}[t_{{\sm 1}}]\right)^{*},\\
\label{eq:diaD3}
D^{pqj}_{lk}(3)&=& \left[\iu(\fr_{pq}+\fr_{jk})+2\epsilon\right]^{-1}\nonumber \\
&&\times\left(\dot{W}^{1}_{jq}W^{1}_{kj}-\dot{W}^{3}_{kp}W^{1}_{kj}\right.\nonumber \\
&&+\left.\dot{W}^{3}_{kp}W^{-1}_{pq}[t_{{\sm 1}}]-\dot{W}^{1}_{jq}W^{1}_{kj}[t_{{\sm 1}}]\right)^{*},\\
\label{eq:diaD8b}
D^{pqj}_{lk}(8-b)&=& \left[\iu(\fr_{jk}+\fr_{lp})+2\epsilon\right]^{-1} \nonumber \\
&&\times\left(\dot{W}^{3}_{jq+lk}\left(W^{1}_{kj}\right)^{*}-\dot{W}^{1}_{pq}\left(W^{1}_{kj}\right)^{*}\right.\nonumber \\
&&+\left.\dot{W}^{3}_{jq+lk}W^{-1}_{kj}[t_{{\sm 1}}]-\dot{W}^{1}_{pq}W^{1}_{lp}[t_{{\sm 1}}]\right),\\
\label{eq:diaD9}
D^{pqj}_{lk}(9)&=& \left[\iu(\fr_{pq}+\fr_{jk})+2\epsilon \right]^{-1}\nonumber \\
&&\times\left(\dot{W}^{1}_{lp}\left(W^{1}_{kj}\right)^{*}-\dot{W}^{3}_{jq+lk}\left(W^{1}_{kj}\right)^{*}\right.\nonumber \\
&&+\left.\dot{W}^{1}_{lp}W^{1}_{pq}[t_{{\sm 1}}]-\dot{W}^{3}_{jq+lk}W^{-1}_{kj}[t_{{\sm 1}}]\right).
\end{eqnarray}
Above some of the $\dot{W}$- and $\dot{W}W[t_{{\sm 1}}]$-functions have subscript labels as $pq+kl$, see first term of $D^{pqj}_{lk}(8-b)$. This notation implies that we should replace $\fr_{pq}$ in Eqs.~(\ref{eq:Wdot}) and (\ref{eq:WWt1}) by $\fr_{pq}+\fr_{kl}$. The $D$-functions are expressed in terms of double integrals after performing one of the integrals analytically as done in case of diagram (1-f).

The remaining diagrams are related to the ones stated above and hence the $4$th order term in the heat current can be expressed as
\begin{eqnarray}
\label{eq:4hcurr}
&&\lambda^4 \left[ \frac{1}{3!} \langle \leftdot{V} V^3 \rangle_{\rho} - 
\frac{1}{2!} \bigl\langle \langle X^T V^2\rangle_\rho \leftdot{V}V \bigr\rangle   
\right]\nonumber \\ 
&&= \lambda^{3}\sum_{\substack{p,q,j \\ k,l}}\rho_{kl}\,u_{pq}\,u_{qj}\,u_{lp}\,u_{jk} \left[D^{pqj}_{lk}-\left(D^{jqp}_{kl}\right)^{*}\right],
\end{eqnarray}
where the tensor $D$ is given by
\begin{equation}
D = D(1-f)+D(2)+D(3)+D(8-b)+D(9).
\end{equation}
In Eq.~(\ref{eq:4hcurr}) the reduced density matrix is the exact one which needs to be replaced by $\rho^{(0)}$ to obtain the current accurate upto $4$th order. The reduced density matrix upto $2$nd order can be obtained using Eqs.~(\ref{eq:qme2}) -- (\ref{eq:qme3}) and diagrams similar to the ones described above by replacing $\leftdot{V}$ by $[X^{T},V]$. Thus, using Eq.~(\ref{eq:2hcurr}) with $\rho = \rho^{(0)}+\lambda^{2}\rho^{(2)}$ and Eq.~(\ref{eq:4hcurr}) with $\rho = \rho^{(0)}$ we can evaluate the heat current in the harmonic oscillator upto $4$th order. It is important to stress here that since the harmonic oscillator has an unbounded spectrum it becomes essential to consider a relatively large system-Hilbert space at high temperatures. This can be easily achieved, in short computational times, using our approach described above due to the simplification from triple to double integrals which drastically reduces the computational complexity of the problem. 

In order to perform the numerics we choose the spectral density of the baths to be of the ohmic form with a Lorentz-Drude cut-off given by
\begin{equation}
\label{eq:specHO}
J_{\alpha}(\omega) = \frac{\eta \gamma \omega}{1+(\omega/\omega_{{\sm D}})^2}.
\end{equation}
Throughout this work we have considered an expansion in the system-bath coupling which ultimately translates into $\eta\gamma$ being small for the spectral density given above. Physically the parameter $\eta\gamma$ being small implies that the relaxation time of the system $\tau_{{\sm R}} \propto [\eta\gamma]^{-1}$ should be much longer than i) the correlation decay time of the baths $\tau_{{\sm B}}$ and ii) the longest time scale of the bare system $\tau_{{\sm S}}$. In case of the ohmic spectral density described above the correlation decay time of the baths $\tau_{{\sm B}} \propto \mathrm{min}\{k_{{\sm B}}T_{{\sm L}}/\hbar, k_{{\sm B}}T_{{\sm R}}/\hbar, \omega_{{\sm D}}\}^{-1}$. Whereas $\tau_{{\sm S}}$ is inversely proportional to the smallest energy difference of the bare system Hamiltonian, which in case of the harmonic oscillator system is given by $\tau_{{\sm S}} \propto \omega_{{\sm 0}}^{-1}$. In other words for the perturbation theory to hold $\eta\gamma$ must satisfy i) $\eta\gamma \ll \mathrm{min}\{k_{{\sm B}}T_{{\sm L}}/\hbar, k_{{\sm B}}T_{{\sm R}}/\hbar, \omega_{{\sm D}}\}$ and ii) $\eta \ll \omega_{{\sm 0}}$. Thus, in terms of the small parameter $\eta$ we split the current into $2$nd order and $4$th order contributions as
\begin{equation}
\label{eq:currspl}
I_{{\sm L}}^{en} = a_{2}\eta + a_{4}\eta^{2}.
\end{equation}
Above $a_{2}$ and $a_{4}$ are the second and fourth order contributions to the current. 

\begin{figure}
 \centering
\includegraphics[width=\columnwidth]{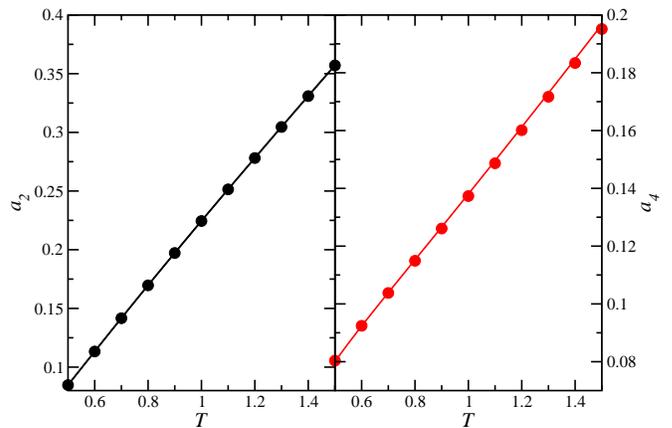}
\caption{\label{fig:HOcomp}(Color online) The coefficients $a_2$ (left panel) and $a_4$ (right panel) for the heat current in the harmonic oscillator model as a function of temperature, $T = (T_{{\sm L}}+T_{{\sm R}})/2$. The temperatures of the left and right bath are set as $T_{{\sm L}} = T (1+\delta T)$ and $T_{{\sm R}} = T (1-\delta T)$ with $\delta T = 0.5$. The harmonic baths are described by an ohmic spectral density with Lorentz-Drude cut-off of $\omega_{{\sm D}} = 1$ and $\gamma = 1$. Solid lines correspond to the exact NEGF results, whereas the dots correspond to the $4$th order improved Dyson series outlined in this work. The system energy scale $\hbar\wz = 1$ and we truncate the energy spectrum of the harmonic oscillator by considering 80 levels. All parameters are in dimensionless units [$k_{{\sm B}} =\hbar = 1$].}
\end{figure}

In Fig.\ref{fig:HOcomp} we compare the results obtained via our improved Dyson series approach to the exact NEGF formalism \cite{wang13review}. The exact $2$nd and $4$th contributions are extracted from the NEGF formalism and plotted as solid lines in the left and right panel of Fig.\ref{fig:HOcomp}. The dots represent the values obtained from our improved Dyson series approach which matches the exact NEGF results remarkably well. 
\subsection{Quantum dot model}
\label{sec:6B}
We now proceed to the QD model described by Eqs.~(\ref{eq:QDC})--(\ref{eq:QDL}). Unlike the harmonic oscillator we can analytically evaluate the electronic current in this model to a great extent relying minimally on the numerics. Thus, we begin with the definition of the electronic current operator
\begin{equation}
\hat{I}_{{\sm L}}^{el} = -\frac{d N_{{\sm L}}}{dt} = -\frac{i}{\hbar}
[H,N_{{\sm L}}] = \lambda \leftdot{V},
\end{equation}
where $H$ is the total Hamiltonian and $N_{{\sm L}}$ is the number operator defined below Eq.~(\ref{eq:QDL}). The anti-Hermitian operator $\leftdot{V}$ is then given by
\begin{eqnarray}
\leftdot{V} &= -\leftdot{V}^\dagger &= \sum_{j}\Bigl( g_{{\sm L},j}^{*}\, c_{{\sm L},j} \, d^\dagger - g_{{\sm L},j}\, c^\dagger_{{\sm L},j}\, d \Bigr).
\end{eqnarray}

\begin{figure}
 \centering
\includegraphics[width=\columnwidth]{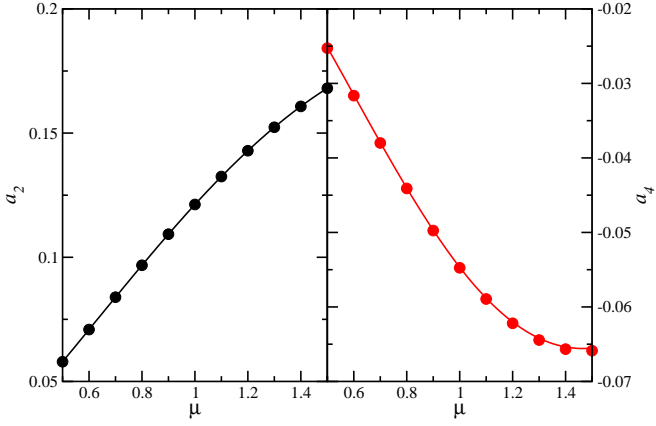}
\caption{\label{fig:QDcomp}(Color online) The coefficients $a_2$ (left panel) and $a_4$ (right panel) for the electronic current in the QD model as a function of chemical potential, $\mu = (\mu_{{\sm L}}+\mu_{{\sm R}})/2$. The chemical potentials of the left and right bath are set as $\mu_{{\sm L}} = \mu (1+\delta\mu)$ and $\mu_{{\sm R}} = \mu (1-\delta\mu)$ with $\delta\mu = 0.5$. The fermionic baths are described by a wide-band spectral density with Lorentz-Drude cut-off of $E_{c} = 10$ and $\Gamma =1$. Solid lines correspond to the exact NEGF results, whereas the dots correspond to the $4$th order improved Dyson series outlined in this work. The system energy scale $E_{{\sm 0}} = 1$ and the temperatures of the baths are set to $T_{{\sm L}} = T_{{\sm R}} = 1$. All parameters are in dimensionless units and $k_{{\sm B}} =\hbar = 1$.}
\end{figure}
In order to evaluate the current upto $4$th order we first define the bath correlators
\begin{eqnarray}
\label{eq:QDCg}
C^{>}_{\alpha}(t) &=& \sum_j | g_{\alpha,j}|^2 {\rm Tr}_{\alpha}\left( \rho_{\alpha}
c_{\alpha,j}(t) c_{\alpha,j}^\dagger(0) \right) \nonumber \\
&=&\int_{-\infty}^{\infty}\frac{dE}{2\pi} \Gamma_{\alpha}(E)\left[1-f_{\alpha}(E)\right]e^{-\frac{\iu}{\hbar} E t},
\end{eqnarray}
\begin{eqnarray}
\label{eq:QDCl}
C^{<}_{\alpha}(t) &=& \sum_j | g_{\alpha,j}|^2 {\rm Tr}_{\alpha}\left( \rho_{\alpha}
c_{\alpha,j}^\dagger(0) c_{\alpha,j} (t) \right)\nonumber \\
&=&\int_{-\infty}^{\infty}\frac{dE}{2\pi} \Gamma_{\alpha}(E)f_{\alpha}(E)e^{-\frac{\iu}{\hbar} E t},
\end{eqnarray}
where $\alpha = L, R$ for the left and right baths respectively, $f_{\alpha}(E) = [e^{\beta_{\alpha}(E-\mu_{\alpha})}+1]^{-1}$ is the Fermi-Dirac distribution of the fermionic baths, and the spectral density $\Gamma_{\alpha}(E) = 2\pi\sum_{j}|g_{\alpha,j}|^{2}\delta(E-\varepsilon_{\alpha,j})$. The associated $W$-functions, which are the Fourier-Laplace transforms of the bath correlators, are defined as
\begin{equation}
\label{eq:QDW}
W^{<,>}(t) = \sum_{\alpha = {\sm L,R}}\int_{-\infty}^t dt'\, e^{\frac{\iu}{\hbar}E_{{\sm 0}} t'} C^{<,>}_{\alpha}(t') e^{\epsilon t'},
\end{equation}
where $E_{{\sm 0}}$ is the energy of the isolated system, see Eq.~(\ref{eq:QDC}). The $C$'s and $W$'s will be of central interest and we will express the steady-state current and the reduced density matrix as functions of these quantities.  

The $2$nd order current using the energy eigenbasis of $H_{{\sm S}}$, as per Eq.~(\ref{eq:4curr}), reads
\begin{eqnarray}
\label{eq:qdotI2}
 \lambda^2 \langle \leftdot{V} V \rangle_{\rho} &=& \lambda^2 \Bigl(  \rho_{{\sm 11}} C_{{\sm L}}^{>}[E_{{\sm 0}}] - \rho_{{\sm 00}} C_{{\sm L}}^{<}[E_{{\sm 0}}] 
\Bigr),
\end{eqnarray}
where $C_{{\sm L}}[E]$ is the Fourier transform of $C_{{\sm L}}(t)$, defined by $C_{{\sm L}}[E] = \int_{-\infty}^{+\infty} C_{{\sm L}}(t) e^{(\iu /\hbar) \,Et} dt$. Above $\rho_{{\sm 00}}$ is the element of the \emph{exact} reduced density matrix in the lower energy state, when no electron is present on the dot. On the other hand $\rho_{{\sm 11}}$ is the element in the higher energy state when the QD is occupied with an electron.

The $4$th order current from Eq.~(\ref{eq:4curr}) can be simplified in terms of $W$'s, using techniques similar to the harmonic oscillator case, as
\begin{widetext} 
\begin{eqnarray}
\label{eq:qdotI4}
&&\lambda^4 \left[ \frac{1}{3!} \langle \leftdot{V} V^3 \rangle_{\rho} - 
\frac{1}{2!} \bigl\langle \langle X^T V^2\rangle_\rho \leftdot{V}V \bigr\rangle   
\right] \nonumber \\
= \frac{2}{\hbar^4} {\rm Re} &\Biggl[& 
\rho_{{\sm 11}} \int_{-\infty}^0\!\!\!dt\, e^{2 \epsilon t} \Bigl[
  \Bigl(W^<(-t) - W^<(0) \Bigr) W_{{\sm L}}^>(t)  
-  W^>(-t) W_{{\sm L}}^<(t)  
-  W^>(0) W_{{\sm L}}^>(t) \Bigr] \nonumber \\
&-&\rho_{{\sm 00}} \int_{-\infty}^0\!\!\!dt\, e^{2 \epsilon t} \Bigl[
 \Bigl(W^>(-t) - W^>(0) \Bigr) W_{{\sm L}}^<(t)
-  W^<(-t) W_{{\sm L}}^>(t)  
-  W^<(0) W_{{\sm L}}^<(t) \Bigr]\Biggr].
\end{eqnarray}
\end{widetext}
Hence for the QD model the off-diagonal elements of the reduced density matrix, $\rho_{{\sm 01}}$ and $\rho_{{\sm 10}}$, are not required to calculate the electronic current.

We evaluate the reduced density matrix upto $2$nd order, i.e., $\rho = \rho^{(0)} +\lambda^2 \rho^{(2)}$ explicitly in Append.\ref{append:C}. This is required to keep the current correct upto the $4$th order, specifically, we should replace $\rho = \rho^{(0)} + \lambda^2 \rho^{(2)}$ in Eq.~(\ref{eq:qdotI2}) and $\rho = \rho^{(0)}$ in Eq.~(\ref{eq:qdotI4}). In order to make numerical calculations we choose the spectral density of the fermionic baths to take the form
\begin{eqnarray}
\label{eq:specQD}
\Gamma_{\alpha}(E)&=&\frac{\eta\Gamma}{1+\left(E/E_{c}\right)^{2}}, 
\end{eqnarray}
where $\alpha = L,R$ and $E_{c}$ is the Lorentz-Drude type cut-off in the wide-band spectral density. In this case the physical weak parameter $\eta\Gamma$ governs the relaxation time of the system $\tau_{{\sm S}} \propto [\eta\Gamma]^{-1}$. Thus, for the quantum dot model the weak parameter $\eta\Gamma$ [see below Eq.~(\ref{eq:specHO})] must satisfy i) $\eta\Gamma \ll \mathrm{min}\{k_{{\sm B}}T_{{\sm L}}, k_{{\sm B}}T_{{\sm R}}, E_{c}\}$ and ii) $\eta\Gamma \ll E_{{\sm 0}}$.

Therefore using Eqs.~(\ref{eq:rho0QD}) and (\ref{eq:rho2QD}) as given in Append.\ref{append:C} we evaluate the electronic current upto $4$th order and compare with the NEGF technique as shown in Fig.\ref{fig:QDcomp}. The left-hand panel shows the second order current coefficient ($a_{2}$), whereas the right-hand panel shows the $4$th order current coefficient ($a_{4}$) split in a similar way to Eq.~(\ref{eq:currspl}). In both panels the dots indicate the results obtained via the approach outlined herein and the solid lines represent the $2$nd and $4$th results extracted from the exact NEGF formalism. Clearly we can see that our approach perfectly matches the NEGF results confirming our method. In case of the QD model if we consider the high-bias limit we can analytically compare the results from our approach to that of NEGF as shown in Append.\ref{append:D}. Thus, the excellent agreement between NEGF and our method for both the harmonic oscillator and the QD model validates our approach beyond reasonable doubt.
\begin{figure}
\includegraphics[width=\columnwidth]{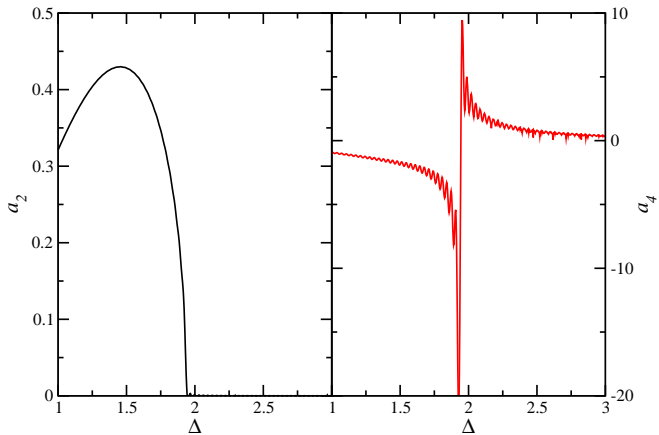}%
\caption{\label{fig:RubinD}(Color online) The coefficients $a_2$ (left panel) and $a_4$ (right panel) for the current expansion as a function of $\Delta$ for the spin-boson model with Rubin baths having $\omega_{{\sm R}}=2$. The other parameters in dimensionless units [$k_{{\sm B}} =\hbar = 1$] are set to $T_{{\sm L}}=1.5$, $T_{{\sm R}}=0.5$, and $E=0.5$.}
\end{figure}
\subsection{Spin-boson model}
\label{sec:6C}
Now we tackle the nonlinear problem of the spin-boson model. This is perhaps the simplest model of a quantum system coupled to an environment. It has been studied extensively in the literature as an archetype model for an atom coupled to an electromagnetic field \cite{leggett87review} in the field of quantum optics.  In recent years, the model is also used to mimic a molecular junction coupled to two baths for thermal transport \cite{segalprl05}. Several nonperturbative approaches, within some underlying assumptions, have been used to study its strong coupling limit and its interesting connection to Kondo problem \cite{saitoprl13, segalpre14}.

In this model, since there is no particle flow, the heat current is same as the energy current. Hence, using the basic definition of energy current we can express the operator $\leftdot{V}$ as
\begin{equation}
\leftdot{V} = \frac{\sigma_z}{2} \iu\hbar\sum_j g_{{\sm L},j} \, P_{{\sm L},j}.
\end{equation} 
The bath correlators $C_{\alpha}(t)$ and $\chi_{\alpha}(t)$ are the correlators for the harmonic oscillator baths defined in Eqs.~(\ref{eq:CoorC}) and (\ref{eq:CoorChi}). Thus, all the formulas for the general harmonic baths can be used without change for the spin-boson model. The only difference is that now there are only two states in the system.

The two-level system Hamiltonian can be diagonalized to give the eigenvalues
$E_{\pm} =\pm\sqrt{E^2 + \Delta^2}\,/2$ and the coupling matrix elements to the baths in the eigenbasis of $H_{{\sm S}}$ are $\sigma_z^{{\sm --}} = -\sigma_z^{{\sm ++}} = \cos\theta$, and 
$\sigma_z^{{\sm +-}} = \sigma_z^{{\sm -+}} = -\sin\theta$ with $\theta = \tan^{-1}(\Delta/E)$. Using the general expression for the second order current one can work out an exact expression \cite{segalprl05, segaljcp05, ruokolaprb11}
\begin{equation}
\label{eq:SBM2I}
a_2 = \frac{2\tilde{\omega} \sin^2\theta J_{{\sm L}}(\tilde{\omega}) J_{{\sm R}}(\tilde{\omega}) \big[ n_{{\sm L}} - n_{{\sm R}}\bigr]}{J_{{\sm L}}(\tilde{\omega})(2 n_{{\sm L}} + 1) +J_{{\sm R}}(\tilde{\omega}) (2 n_{{\sm R}} + 1) }, 
\end{equation}
where $J_{\alpha}(\tilde{\omega})$ is the bath spectral density and the Bose-Einstein distribution function $n_{{\sm L}}$ and $n_{{\sm R}}$ are evaluated at $\tilde{\omega} = \sqrt{E^2 + \Delta^2}\,/\hbar$. The $4$th order coefficient $a_4$ is analytically cumbersome and hence in this work we determine it numerically. 

In Fig.\ref{fig:RubinD} we present $a_2$ and $a_4$ for the Rubin bath with a spectral density, $J_{\alpha}(\omega) = (\hbar \eta\omega/2) \sqrt{\omega_{{\sm R}}^{2} - \omega^2}\,\Theta(\omega_{{\sm R}}-\omega)$ with $\alpha = L,R$. A striking feature is that $a_4$ changes sign when the system energy spacing $\hbar\tilde{\omega}$ is larger than the band width of the bosonic baths. There is a sort of resonance exactly at the band edge. Since $a_2$ is $0$ above the phonon band, $a_4 > 0$ can be interpreted as a two-phonon transmission (or a cotunnelling) process. We also note that for small values of $\Delta$, before crossing over to cotunnelling regime, large coupling suppresses the current.

\begin{figure}
\includegraphics[width=\columnwidth]{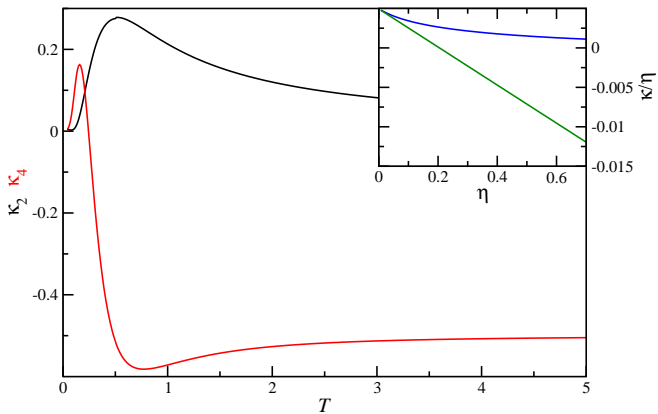}%
\caption{\label{fig:LRkappa}(Color online) The coefficients $\kappa_2$ (black online) and $\kappa_4$ (red online) for the thermal conductance of a spin-boson model as a function of temperature $T$ for $E=0$ and $\Delta=1$. The inset shows the comparison between our approach (green online) and the work of Segal (ref.~[\onlinecite{segalpre14}]) (blue online) for the conductance $\kappa$ at $T = 1$, $E=0$, and $\Delta = 0.1$. The baths for both plots are of Lorentz-Drude type with $\omega_{{\sm D}}=1$. All quantities are expressed in dimensionless units by setting $k_{{\sm B}} =\hbar = 1$.}
\end{figure}
Similar phenomenon shows up in the Lorentz-Drude model with a spectral density, $J_{\alpha}(\omega) = \eta \hbar\omega [1+(\omega/\omega_{{\sm D}})^2]^{-1}$ with $\alpha = L,R$, in a different way.  In Fig.\ref{fig:LRkappa} we plot the conductance $\kappa = dI_{{\sm L}}^{en}/dT = \kappa_2 \eta + \kappa_4 \eta^2$ determined numerically. In this case, $\kappa_4$ is positive at low temperatures. Nonperturbative analysis\cite{ruokolaprb11, saitoprl13} shows that for finite $\eta$ the low-temperature asymptotic behaviour is proportional to $\eta T^3$ instead of the exponential suppression as given by $a_2$, see Eq.~(\ref{eq:SBM2I}). This positive $\kappa_4$ strongly indicates an attempt to reach the $T^{3}$ behaviour, from the exponential suppression, in presence of strong system-bath coupling. However, since our analysis is based on a perturbation in the system-bath coupling we must warn that in the low-temperature regime the perturbative theories should generally fail since the higher order coupling strengths become more important in this regime. On the other hand, in the high-temperature regime, the perturbative results are quite reliable. The inset in Fig.\ref{fig:LRkappa} shows comparison between our approach (black line) and the work of Segal \cite{segalpre14} (blue curve), which employs a perturbative analysis valid in the regime of small $\Delta$. Even at small values of $\eta$ (see close to $\eta = 0.05$) the contribution from the $4$th order term, $\kappa_4$, becomes significant to give the correct slope matching well with the work of Segal. As expected our theory fails at large values of $\eta$ giving rise to unphysical negative thermal conductance.
\section{Concluding remarks}
\label{sec:7}
In summary, we have presented a general formulation to evaluate currents in nonequilibrium steady states based on the Dyson expansion approach. We have explicitly shown that the steady-state divergences in the truncated-Dyson series is rooted in the time-translational invariance of the system, which can be resolved by an appropriate inverse mapping of the reduced density matrix at long but \emph{finite} times resulting in a suitable choice of initial condition. The improved Dyson series, based on this prudent choice, outlined in this work could be used for \emph{any} general nonlinear system Hamiltonians, e.g. quantum dot models with electron-electron interaction or nanoelectromechanical systems with electron-vibration interaction, and is equivalent to solving the full master equation order by order. The improved Dyson series not only circumvents the divergences, yielding finite results in the steady state, but also ensures that the steady-state results are independent of the initial conditions. It also does not make any \emph{a priori} assumptions on the type of coupling between the baths and the system giving it an advantage over the path-integral approaches, where the position operator of the bath typically couples to the system Hamiltonian. 

As an application of our formalism we evaluated heat and electronic currents upto $4$th order in system-bath coupling, an improvement over the weak-coupling theories based on quantum master equations. We verified our approach for the case of noninteracting bosons and fermions with the nonequilibrium Green's function method and found remarkable agreement. We also evaluated heat currents in the nonlinear spin-boson model and found strong evidence signalling towards cotunnelling process.

Overall, our approach provides a systematic way to evaluate currents in general nonlinear multi-level systems well beyond the capability of exact simulations which are limited to a few levels. Even though our approach provides a possible route to go beyond the weak-coupling regime several open questions still persist, like the extension of our method to higher orders of system-bath coupling, possibly using Monte-Carlo techniques. Another interesting avenue would be to study the asymptotic effects of time-periodic forces on currents using the Floquet basis \cite{grifoniphysrep98}. These fascinating future aspects would help develop a general formalism to understand the transport properties of nonlinear molecular junctions beyond the weak system-bath coupling regime.  

\section*{Acknowledgments}
We would like to thank Peter H\"{a}nggi for useful remarks.

\appendix
\section{Identity for linear baths}
\label{append:A}
In this appendix we prove a general identity [Eq.~(\ref{eq:xv-eps}) from the main text] using which we can show that each term of the truncated-Dyson series diverges in the steady-state limit. In order to prove this relation we start as per the real-time definition in Eq.~(\ref{eq:XVn}) as
\begin{eqnarray}
\frac{1}{n!} \langle X^{T} V^n \rangle_{\rho_d} &=& \int_{\times n} \mathrm{Tr} \left[ X^{T}(0) [V(t_1),\cdots , V(t_n),\rho_{d}]\right] \nonumber \\
&=& \int_{\times n} \mathrm{Tr} \left[ \rho_{d} [\![X^{T}(0), V(t_1),\cdots , V(t_n)]\!]\right],\nonumber
\end{eqnarray}
where we have used the short-hand notation $\int_{\times n} = \int_{-\infty}^{0}dt_{1}\int_{-\infty}^{t_{1}}dt_{2}\cdots\int_{-\infty}^{t_{n-1}}dt_{n}$ and the initial time $t_{{\sm 0}} = -\infty$. In the first line above we have used the right normed convention for the nested commutators which has been transformed into the left normed convention using the cyclic property of trace, i.e., $\mathrm{Tr}\left([A,B]C\right) = \mathrm{Tr}\left(A[B,C]\right)$, for the second line. The left normed convention is defined as, e.g., $[\![X^{T}(0),V(t_1),V(t_2),V(t_3)]\!] \equiv [[[X^{T}(0),V(t_1)],V(t_2)],V(t_3)]$. The operators within the trace are in the interaction picture, i.e., $V(t) = e^{(\iu /\hbar)\,\h t} V_{\Sc} e^{-(\iu /\hbar)\, \h t}\,e^{\epsilon t}$, where the operator $V_{\Sc}$ is in the Schr\"odinger picture and $\epsilon$ is the adiabatic switch-on parameter. Noting that $[e^{(\iu /\hbar)\, \h t},\rho_{d}] = 0$ (for any time $t$), where $\rho_{d}$ is diagonal in the eigenbasis of $H_{{\sm S}}$, and inserting the identity operator $\mathds{1} = e^{-(\iu /\hbar)\, \h t_{1}}e^{(\iu /\hbar)\, \h t_{1}}$ appropriately we get
\begin{eqnarray}
& &\frac{1}{n!} \langle X^{T} V^n \rangle_{\rho_d} = \int_{\times n}e^{n\epsilon t_{1}} \nonumber \\ 
& &\times \mathrm{Tr} \left[ \rho_{d} [\![X^{T}(-t_{1}), V, V(t_{2}-t_{1}),\cdots , V(t_{n}-t_{1})]\!]\right].\nonumber 
\end{eqnarray}
Making a change of variables in the time integration as $t_{2}-t_{1} = t_{1}^{\prime},\, t_{3}-t_{1} = t_{2}^{\prime}, \cdots , t_{n}-t_{1} = t_{n-1}^{\prime}$ and noting that $X_{mn} = |m \rangle\langle n|$ we can analytically perform the $t_{1}$ integral to obtain our final expression
\begin{eqnarray}
\label{eq:IdB}
\frac{1}{n} \langle X^{T} V^n \rangle_{\rho_d} = 
\frac{1}{\iu \fr + n \epsilon} \langle [X^{T}, V]V^{n-1} \rangle_{\rho_d}.
\end{eqnarray}
Above the super-operator $\fr$ must be interpreted carefully. The operator $\fr = (E_{m}-E_{n})/\hbar$ if and only if the operator $X = |m\rangle\langle n|$. In other words the operator $\fr$ can be interpreted only in the basis of the system Hamiltonian $H_{{\sm S}}$ and it takes energy difference values, $(E_{m}-E_{n})/\hbar$, corresponding to the Hubbard operator which transforms the state $|n\rangle$ to the state $|m\rangle$. In matrix element form Eq.~(\ref{eq:IdB}) can be expressed as
\begin{eqnarray}
\frac{1}{k} \langle |n\rangle\langle m| V^k \rangle_{\rho_d} = 
\frac{1}{\iu \fr_{mn} + k \epsilon} \langle [|n\rangle\langle m|, V]V^{k-1} \rangle_{\rho_d},\nonumber
\end{eqnarray}
where $\fr_{mn} = (E_{m}-E_{n})/\hbar$.
\section{Reduced density matrix for the quantum dot model}
\label{append:C}

Here we work out the reduced density matrix for the QD model upto $2$nd order in system-bath coupling. This requires us to solve a $4$th order QME which is well beyond the standard weak-coupling approach. We begin by expressing each term of the $4$th order QME, Eq.~(\ref{eq:4qme}), in terms of $W$'s and the Fourier transforms of $C$'s, Eqs.~(\ref{eq:QDCg})--(\ref{eq:QDW}). The trivial $0$th order term can then be expressed as
\begin{equation}
-\frac{\iu}{\hbar} [H_{{\sm S}},\rho] = \left(   \begin{array}{cc}
0 & \frac{\iu}{\hbar} E_{{\sm 0}} \rho_{{\sm 01}} \\
-\frac{\iu}{\hbar} E_{{\sm 0}} \rho_{{\sm 10}} & 0 
\end{array}
\right).
\end{equation}
Above we have used the system's eigenenergy basis to decompose the \emph{exact} reduced density matrix into diagonal ($\rho_{{\sm 00}}$ and $\rho_{{\sm 11}}$) and off-diagonal ($\rho_{{\sm 01}}$ and $\rho_{{\sm 10}}$) terms. The state $|0\rangle$ corresponds to the state when the QD has no electron present on it, i.e., the lower-energy state and the state $|1\rangle$ corresponds to the higher energy state when an electron is present on the QD.

The $2$nd order term of the QME reads
\begin{eqnarray}
\label{eq:QD2rho}
\lambda^2 \langle [X^T,V]V \rangle_{\rho} = -\frac{1}{\hbar^2} 
\left( \begin{array}{cc}
A_{{\sm 00}} & A_{{\sm 01}}  \\
A_{{\sm 10}} & A_{{\sm 11}}  
\end{array}
\right),
\end{eqnarray}
where the elements of the matrix $A$ are given by,
\begin{eqnarray}
\label{eq:A00}
A_{{\sm 00}} &= -A_{{\sm 11}} &= \rho_{{\sm 00}} C^<[E_{{\sm 0}}]-\rho_{{\sm 11}} C^>[E_{{\sm 0}}], \\
\label{eq:A01}
A_{{\sm 01}} &= \left(A_{{\sm 10}}\right)^{*} &= \rho_{{\sm 01}} \bigl(W^<(0) +W^>(0)\bigr).
\end{eqnarray}
Above $C^{<,>}[E_{{\sm 0}}] = C^{<,>}_{{\sm L}}[E_{{\sm 0}}] + C^{<,>}_{{\sm R}}[E_{{\sm 0}}]$.

Similarly the $4$th order terms can be written as
\begin{eqnarray}
\label{eq:QD4rho}
\frac{\lambda^4}{3!} \langle [X^T,V]V^3 \rangle_{\rho} 
&-& \frac{\lambda^4}{2!}\bigl\langle \langle X^TV^2\rangle_{\rho} [X^T,V] V \bigr\rangle \nonumber \\ &=&\frac{1}{\hbar^4} 
\left( \begin{array}{cc}
B_{{\sm 00}} & B_{{\sm 01}}  \\
B_{{\sm 10}} & B_{{\sm 11}}  
\end{array}
\right),
\end{eqnarray}
where the elements of the matrix $B$ are given by
\begin{eqnarray}
\label{eq:B00}
B_{{\sm 00}} &=& - B_{{\sm 11}}  \\
&=& -\rho_{{\sm 00}} 
\int_{-\infty}^0 dt\, e^{2 \epsilon t}  \Big[ W^<(t)\bigl(W^{<}(0) + W^{>}(0) \bigr) \nonumber \\
&&-W^>(-t)W^<(t) + W^{<}(-t) W^{>}(t) + {\rm c.c.} \Big] \nonumber \\
&&+ \rho_{{\sm 11}} 
\int_{-\infty}^0 dt\, e^{2 \epsilon t}  \Big[ W^>(t)\bigl(W^{<}(0) + W^{>}(0) \bigr) \nonumber \\
&&-W^<(-t)W^>(t) + W^{>}(-t) W^{<}(t) + {\rm c.c.} \Big], \nonumber \\
\label{eq:B01}
B_{{\sm 01}} &=& B_{{\sm 10}}^{*} \\
&=& 
\rho_{{\sm 01}} 
\int_{-\infty}^0 dt\, e^{2 \epsilon t}  \Big[ -W^<(t)W^{<}(0)  \nonumber \\
&&-W^>(t)W^{>}(0) -W^>(-t)W^<(0) \nonumber \\ 
&&+ W^{>}(-t) W^{<}(t) + W^{<}(-t) W^{>}(t) \Big]. \nonumber
\end{eqnarray}
Above ${\rm c.c}$ stands for complex conjugate.

Now given all the terms of the $4$th order QME explicitly in terms of $W$'s and $C$'s we can easily construct the $0$th and $2$nd order solutions to the QME using the order-by-order method described in Sec.\ref{sec:5}. Using Eqs.~(\ref{eq:qme2}), (\ref{eq:qme1}), and (\ref{eq:QD2rho}) and imposing the normalization condition $\mathrm{Tr}(\rho^{(0)}) = 1$ we obtain the $0$th order solution as
\begin{eqnarray}
\label{eq:rho0QD}
\rho^{(0)} =\left( \begin{array}{cc}
\frac{C^>[E_{{\sm 0}}]}{C^>[E_{{\sm 0}}]+C^<[E_{{\sm 0}}]} & 0 \\
0 & \frac{C^<[E_{{\sm 0}}]}{C^>[E_{{\sm 0}}]+C^<[E_{{\sm 0}}]} 
\end{array}
\right). 
\end{eqnarray}

The $2$nd order solution can be obtained using Eqs.~(\ref{eq:qme4}),~(\ref{eq:qme3}), and~(\ref{eq:QD4rho}) as
\begin{eqnarray}
\label{eq:rho2QD}
\rho^{(2)} =\left( \begin{array}{cc}
-\frac{ B_{{\sm 00}}^{(0)}}{C^>[E_{{\sm 0}}] + C^<[E_{{\sm 0}}] } & 0 \\
0 & \frac{ B_{{\sm 00}}^{(0)}}{C^>[E_{{\sm 0}}] + C^<[E_{{\sm 0}}]} 
\end{array}
\right), 
\end{eqnarray}
where $B_{{\sm 00}}^{(0)}$ is obtained by replacing $\rho_{{\sm 00}}$ by $\rho_{{\sm 00}}^{(0)}$ and $\rho_{{\sm 11}}$ by $\rho_{{\sm 11}}^{(0)}$ in Eq.~(\ref{eq:B00}). Thus Eqs.~(\ref{eq:rho0QD}) and~(\ref{eq:rho2QD}) form the steady-state solution of the $4$th order QME correct upto $2$nd order in the system-bath coupling.
 
\section{High-bias limit for the quantum dot model}
\label{append:D}
In order to calculate current in the high bias limit we take $\mu_L\rightarrow+\infty$ and $\mu_R\rightarrow-\infty$. In this limit the exact expansion of current with respect to the system-bath coupling can be obtained from the NEGF formalism. Hence in this appendix we show that our $2$nd and $4$th order current formalism analytically match the exact expressions from NEGF.
	
In the high bias limit the Fermi-Dirac distribution is a constant, such that $f_L(E)=1$ and $f_R(E)=0$, for all energies $E$. Therefore the bath correlators for the wide-band spectral density with the Lorentz-Drude cut-off, Eq.~(\ref{eq:specQD}), can be evaluated as
\begin{equation}
C_{{\sm L}}^<(t)=C_{{\sm R}}^>(t)=\frac{\eta\Gamma E_{c}}{2\hbar} e^{-\frac{E_{c}}{\hbar}|t|},
\end{equation}
while $C_{{\sm L}}^>(t)=C_{{\sm R}}^<(t)=0$. Hence, the Fourier transform of the correlators can be easily evaluated as
\begin{equation}
C^<_{{\sm L}}[E_{{\sm 0}}]=C^>_{{\sm R}}[E_{{\sm 0}}]=-\frac{\eta\Gamma\hbar^2 E_{c}^2}{E_{{\sm 0}}^2 + E_{c}^2}.
\end{equation}
Then the $0$th order reduced density matrix $\rho^{(0)}$ according to Eq.~(\ref{eq:rho0QD}) reads
\begin{equation}
\rho^{(0)}_{{\sm 00}}=\rho^{(0)}_{{\sm 11}}=\frac{1}{2},
\end{equation}
while the off-diagonal elements are exactly zero. Subsequently the $2$nd order current can be calculated according to Eq.~(\ref{eq:qdotI2}) as
\begin{equation}
\label{eq:QDI2HB}
\eta a_{2}=\lambda^2 \langle \leftdot{V} V \rangle_{\rho^{(0)}} = \frac{\eta \Gamma E_{c}^2}{2\left(E_{{\sm 0}}^{2}+E_{c}^{2}\right)}.
\end{equation}

In order to evaluate the $4$th order current we require the $W$-functions, defined by Eq.~(\ref{eq:QDW}), which can be obtained in the high-bias limit for $t< 0$ as
\begin{equation}
W_{{\sm L}}^<(t) = \frac{\eta \hbar\Gamma E_{c}}{2}\left(\frac{2E_{c}}{E_{{\sm 0}}^{2}+E_{c}^2}-\frac{e^{-\frac{t}{\hbar}(E_{c}-\iu E_{{\sm 0}})}}{E_{c}-\iu E_{{\sm 0}}}\right)
\end{equation}
while $W_{{\sm R}}^>(t)=W_{{\sm L}}^<(t)$ and $W_{{\sm L}}^>(t)=W_{{\sm R}}^<(t)=0$. With these $W$-functions and using Eq.~(\ref{eq:rho2QD}) one can immediately show that the $2$nd order reduced density matrix is exactly zero, i.e., $\rho^{(2)}=0$. Hence Eq.~(\ref{eq:qdotI4}) becomes the exact formula for the $4$th order current, which can be simplified as
\begin{eqnarray}
\eta^{2} a_{4} &=& \lambda^4 \left[ \frac{1}{3!} \langle \leftdot{V} V^3 \rangle_{\rho^{(0)}} - 
\frac{1}{2!} \bigl\langle \langle X^T V^2\rangle_{\rho^{(0)}} \leftdot{V}V \bigr\rangle   
\right] \nonumber \\
&=&\frac{1}{\hbar^4}\mbox{Re}\Biggl[\int_{-\infty}^0dte^{2\epsilon t}\Bigl[W_L^<(0)W^<_L(t)\nonumber\\
&&+W_R^>(0)W^<_L(t)-2W_R^>(-t)W_L^<(t)\Bigr]\Biggr].
\end{eqnarray}
The above integrations can be carried out analytically and one gets
\begin{equation}
\label{eq:QDI4HB}
\eta^{2} a_{4} = \frac{\eta^2 \Gamma^{2} E_{c}(E_{{\sm 0}}^2-E_{c}^2)}{4\hbar(E_{{\sm 0}}^{2}+E_{c}^{2})^2}.
\end{equation}
This is the final formula of the fourth order current, where the $\epsilon\rightarrow 0$ limit has already been taken. 

For analytical comparison, we now evaluate the current from the Landauer formula in the high bias limit
\begin{eqnarray}
I_{{\sm L}}^{el}&=&\int_{-\infty}^{+\infty}\frac{dE}{\hbar}\frac{\eta^2 \Gamma^{2}E_{c}^4}{\left[(E-E_{{\sm 0}})(E^2+E_{c}^2)-\eta\Gamma E E_{c}\right]^2+E_{c}^4\eta^2\Gamma^{2}}\nonumber\\
&=&\frac{E_{c}(\eta\Gamma+2E_{c})}{\hbar\left[(\eta\Gamma+2E_{c})^2+4E_{{\sm 0}}^2\right]}.
\end{eqnarray}
Thus, one can now expand $I_{{\sm L}}^{el}$ with respect to $\eta$ to obtain the series expansion
\begin{equation}
I_{{\sm L}}^{el}=-\frac{E_{c}}{\hbar}\sum_{n=1}^\infty\mathrm{Re}\left[\left(\frac{x}{2}\right)^n\right]\eta^n ,
\end{equation}
where $x=-\Gamma(E_{c}-\iu E_{{\sm 0}})/(E_{{\sm 0}}^2+E_{c}^2)$. Thus, it can be clearly seen that the first two terms of the above series correspond to the improved Dyson series results, i.e., $n=1$ corresponds to the $2$nd order result Eq.~(\ref{eq:QDI2HB}), whereas $n=2$ corresponds to the $4$th order result Eq.~(\ref{eq:QDI4HB}).

\bibliography{dyson-bib}

\end{document}